\documentclass[fleqn,usenatbib]{mnras}

\usepackage{newtxtext,newtxmath}

\usepackage[T1]{fontenc}
\usepackage{ae,aecompl}

\usepackage{setspace}

\usepackage{graphicx}	
\usepackage{multicol}
\usepackage{rotating}
\usepackage{afterpage}
\usepackage{siunitx}
\usepackage{float}
\usepackage{bm}
\usepackage{booktabs,caption}
\usepackage[flushleft]{threeparttable}
\usepackage{hyperref}
\usepackage{pdflscape}
\usepackage{afterpage}
\usepackage{mwe}
\usepackage{dcolumn}
\newcolumntype{d}[1]{D{.}{.}{#1} }
\usepackage{tabularx}
\usepackage{amsmath}	
\usepackage{amssymb}	
\usepackage{color}

\def\fm{\hbox{$.\!\!^{\rm m}$}} 
\def\mnras{MNRAS}
\def\aap{A\&A}
\def\apj{ApJ}
\def\apjs{ApJS}  
\def\aj{AJ}

\def\pasa{PASA}
\def\nat{Nature}

\title[Galaxy Groups in the Completed 2MRS]{The 2MASS redshift survey galaxy group catalogue derived from a graph-theory based friends-of-friends algorithm}

\author[T. S. Lambert et al.]{
Trystan S. Lambert$^{1,2}$ \thanks{E-mail: TrystanScottLambert@gmail.com},
R. C. Kraan-Korteweg$^{2}$,
T. H. Jarrett$^{2}$
and L. M. Macri$^{3}$
\\
$^{1}$ South African Astronomical Observatory, PO Box 9, Observatory 7935, Cape Town, South Africa\\
$^{2}$ Department of Astronomy, University of Cape Town, Private Bag X3,Rondebosch 7701, South Africa\\
$^{3}$ George P. and Cynthia Woods Mitchell Institute for Fundamental Physics and Astronomy,\\Department of Physics and Astronomy, Texas A$\&$M University, 4242 TAMU, College Station, TX 77843, USA.
}

\date{Accepted 2020 June 30. Received 2020 June 30; in original form 2020 February 27}
\pubyear{2020}

\begin{document}
\label{firstpage}
\pagerange{\pageref{firstpage}--\pageref{lastpage}}
\maketitle


\begin{abstract}
 We present the galaxy group catalogue for the recently-completed 2MASS Redshift Survey \citep[2MRS,][]{Macri2019} which consists of 44572 redshifts, including 1041 new measurements for galaxies mostly located within the Zone of Avoidance.
 The galaxy group catalogue is generated by using a novel, graph-theory based, modified version of the Friends-of-Friends algorithm. Several graph-theory examples are presented throughout this paper, including a new method for identifying substructures within groups. The results and graph-theory methods have been thoroughly interrogated against previous 2MRS group catalogues and a Theoretical Astrophysical Observatory (TAO) mock by making use of cutting-edge visualization techniques including immersive facilities, a digital planetarium, and virtual reality. This has resulted in a stable and robust catalogue with on-sky positions and line-of-sight distances within 0.5~Mpc and 2~Mpc, respectively, and has recovered all major groups and clusters. The final catalogue consists of 3022 groups, resulting in the most complete ``whole-sky'' galaxy group catalogue to date. We determine the 3D positions of these groups, as well as their luminosity and comoving distances, observed and corrected number of members, richness metric, velocity dispersion, and estimates of $R_{200}$ and $M_{200}$. We present three additional data products, i.e. the 2MRS galaxies found in groups, a catalogue of subgroups, and a catalogue of 687 new group candidates with no counterparts in previous 2MRS-based analyses.  
\end{abstract}


\section{Introduction}

Galaxies are not isolated in space. They form part of larger structures such as groups, clusters, filaments and voids that together compose the so-called cosmic web \citep{Davis1982,Zeldovich1982,Klypin1993,Bond1996,Jarrett2004}. These large-scale structures are immediately revealed when visually examining the distributions of galaxy positions in redshift space \citep{Peebles1980,Huchra1982,TullyBook1987,Peebles1993Book,FairallBook1998,Huchra2005,Huchra2007,Robotham2011,Huchra2012,Tempel2016,Tempel2018}. However, visually identifying and classifying structures is a subjective method that is neither robust nor quantitative \citep{Huchra2007,Robotham2011,Huchra2012,Mehmet2014,Saulder2016,Jarrett2017,Kourkchi}. As a result, several algorithmic and computational methods for identifying large-scale structure have been put forward. One of the most established and applied methods over the past decades has been the Friends-of-Friends (FoF) algorithm \citep{Huchra1982}, which has been extensively used to find galaxy groups in numerous magnitude-limited redshift surveys \citep{Huchra1982,Ramella1997,TB1997,Huchra2007,Robotham2011}.  

Redshift surveys provide one of the most convenient and efficient tools to study the three-dimensional distribution of galaxies. Being able to objectively identify large-scale structures within these surveys is relevant to addressing a number of unanswered questions in observational cosmology related to the CMB dipole anisotropy \citep{Rauzy1998,Bilicki2011}, establishing a relationship between galaxies and dark matter halos and probing dark matter interactions \citep{Coutinho2016,Lu2016}, constraining cosmological models \citep{Peebles1980},  and exploring the effect of environment on galaxy evolution \citep{Dressler1980,Tempel2016,Tempel2018}. Redshift surveys come in a variety of flavors; the two most common ones are those constrained to a small solid angle on the sky and targeting galaxies at high redshifts (so-called ``pencil beam'' surveys), and those covering larger fractions of the sky at the expense of depth. The exact nature of the redshift survey depends on the science goals at hand, instrumentation, and availability of telescope time \citep{FairallBook1998}. Although numerous large-area redshift surveys have been made over the decades, none span the entire 4$\pi$~sr of the sky. The overwhelming majority of so-called ``whole-sky'' surveys exclude the regions close to the plane of the Milky Way where high stellar densities, an increasingly brighter background, and high levels of extinction make observing galaxies very difficult \citep{Saunders2000,KK2018}. This region of exclusion is often referred to as the Zone of Avoidance (ZoA).
 
As a result of the ZoA, a truly ``whole-sky'' redshift survey does not exist. The requirement for such a survey has been highlighted for several decades to fully explain the CMB dipole anisotropy, a well-known signature caused by the peculiar motion of the Milky Way \citep{LineWeaver1996,Rauzy1998}. This motion cannot adequately be accounted for due to the obscuration by the ZoA of large mass concentrations such as the Great Attractor \citep{Lynden-Bell1988,kk2000} and the recently-discovered Vela Supercluster \citep{KK2015,kk2017,Courtois2019}. While a truly whole-sky magnitude-limited redshift survey does not exist, imaging surveys have been made of the entire sky. Notably among these is the 2 Micron All Sky Survey (2MASS), which uniformly mapped  99.998$\%$ of the sky in the $J$, $H$, and $K_{\rm s}$ bands. Near-infrared light is far less affected by dust in the Galactic plane, resulting in a less-prominent ZoA in 2MASS. The 2MASS extended source catalogue \citep[2MASX,][]{Jarrett2004,Strutskie2006} contains $\sim1.6$~million sources with $K_{\rm s}^{\rm o}< 13 \fm 5$ and is the most complete all-sky galaxy catalogue to date. However, the sheer number density of stars in the Galactic plane makes 2MASX incomplete below $|b|\sim 5-8^\circ$ depending on Galactic longitude \citep{Jarrett2000}.

\begin{figure}
    \centering
     \includegraphics[width=0.5\textwidth]{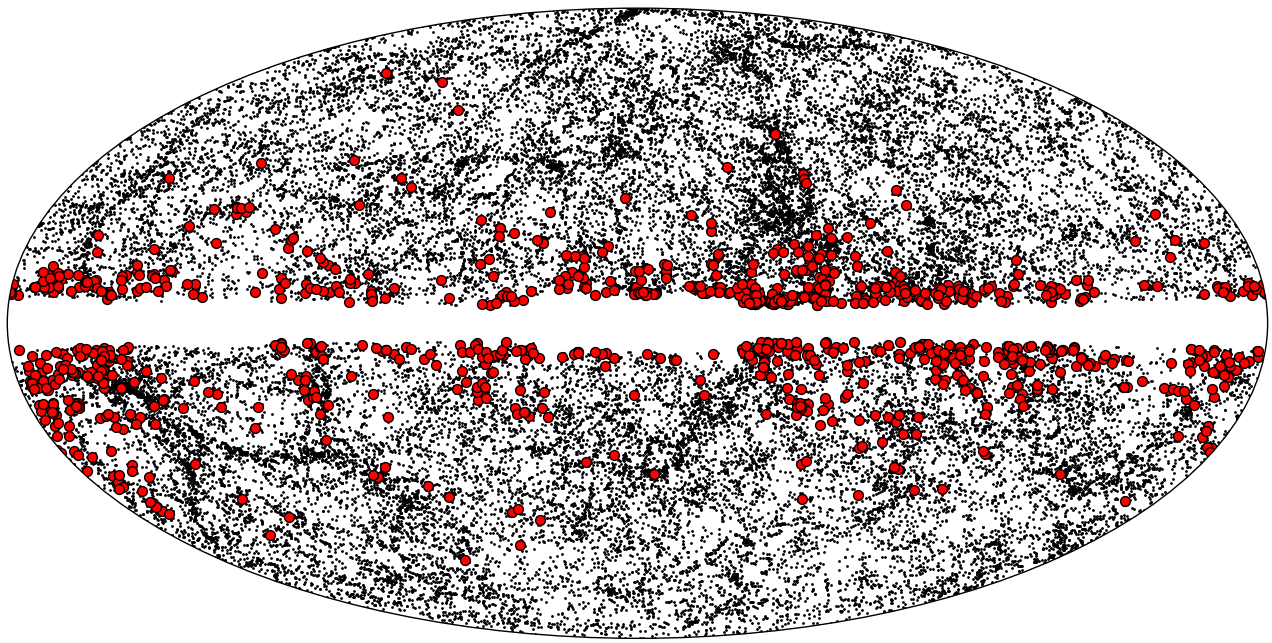}
     \caption{Aitoff projection in Galactic coordinates of the complete 2MRS. Black and red dots represent the galaxies with redshifts from \citet{Huchra2012} and \citet{Macri2019}, respectively. The latter have been scaled for visibility.}
     \label{fig:2MRS2012}
 \end{figure}

The whole-sky nature (and diminished ZoA impact) of 2MASX enabled the 2MASS Redshift Survey \citep[2MRS;][]{Huchra2005,Huchra2012,Macri2019}, a twenty-year concerted effort to measure the redshifts of all $\sim 45000$~2MASX sources with $K^{\rm o}_{\rm s} < 11\fm75$ and thus obtain the most comprehensive (coverage-wise) whole-sky redshift survey to date. A preliminary data release with a magnitude limit of $K^{\rm o}_{\rm s} < 11\fm25$ consisted of $\sim24000$ galaxies \citep{Huchra2005}. The initial data release that reached the target magnitude depth contained over 44000 galaxies \citep{Huchra2012}, but was still significantly incomplete at low Galactic latitudes; the final data release \citep{Macri2019} added the remaining $\sim 1000$ redshifts, mostly in the ZoA. Figure~\ref{fig:2MRS2012} shows the distribution of measurements from the two latter papers. The recent completion of 2MRS, with most of the new redshifts lying close to the ZoA, allows for new large-scale structures in this thinly-mapped area to be uncovered.
 
 A further benefit of having a complete whole-sky galaxy catalogue covering the local Universe is that it lends itself well to the derivation of a group catalogue, one that is more complete along the ZoA than previous work. These catalogues can be complementary to dedicated surveys within the ZoA \citep[e.g.,][]{KK2016Parkes, Ramatsoku2016,KK2018}. A complete whole-sky local-universe group catalogue is also required to quantify other large-scale structures such as filaments and voids \citep{Mehmet2014}, which are useful in conjunction with other studies of galaxy flows due to large-scale structures in the local Universe \citep{Pomarade2017,Courtois2019,Tully2019}. Furthermore, a complete 2MRS group catalogue based on spectroscopic redshifts can be compared to one based on photometric redshift estimates \citep[2MPZ,][]{Bilicki2013}, which may improve the accuracy of the latter technique. 
 
  In order to characterize the large-scale structure revealed by these new redshifts we have modified the FoF algorithm and, after optimisation, applied it to the final (deepest and complete) version of 2MRS. This modified version of the FoF algorithm aims to improve upon the already successful traditional one by addressing several shortcomings such as sensitivity to initial conditions, non-generalizable parameter selection, and non-physical group membership identification. In this paper we present a FoF-based group finder resulting in what we believe to be a highly accurate and robust ``whole-sky'' galaxy group catalogue. We verified our results by comparing them against previous galaxy catalogues based on earlier versions of 2MRS and a deep mock catalogue to test for various (Malmquist-like) biases in the algorithm, and by performing an exhaustive visual inspection using state-of-the art 3-D visualization tools.   
 
 Galaxies are not homogeneous in their physical make-up. They come in numerous shapes, sizes, masses, morphologies, and chemical make up \citep{Buta2011,Jarrett2017}. How galaxies form, evolve, build their stellar populations and form their distinct morphologies presents many challenges to our understanding of the universe. It is widely agreed that the effect of environment on the make up of a galaxy is significant \citep{Dressler1980,Cluver2018,Bianconi2020,Carlesi2020,Otter2020}. Studying the role that environment has on galaxy formation and evolution requires knowledge not only of the positions of the galaxies (which act as tracers of the baryonic mass throughout the cosmic web) but also the environments in which these galaxies find themselves. Thus, redshift surveys and their corresponding galaxy group catalogues make for very important tools in decoding galaxy formation and evolution \citep{Dressler1980,FairallBook1998,Mehmet2014,Jarrett2017}.
 
 As has been previously mentioned there are two main types of redshift surveys, namely very narrow and deep surveys and wide and shallow surveys. Narrow surveys are excellent tools for studying galaxy evolution and formation through cosmic time while shallow surveys lend themselves very well to exploring the role of environment on galaxy morphology in the Local Universe ($z < 0.2$). Nearby galaxies in the local universe are most often found in galaxy groups \citep{VandeWeygaert2008,Robotham2011,Cluver2018}. Thus, the 2MRS galaxy group catalogue (being the widest area survey to date) makes for an optimal survey to explore the effects of galaxy environments ranging from small groups (e.g., Local Group) to the largest of clusters (e.g., Coma Cluster);  previous 2MRS group catalogues have been successfully applied to environmental studies of galaxies \cite[e.g.,][]{Obrien2018,Calderon2019,Greene2019}.
 
The remainder of this paper is laid out as follows: we describe our method of identifying groups in the 2MRS in \S2, including the modifications we have made to the FoF algorithm. The numerous visualization techniques which were used to interrogate our methods are discussed in \S3. The final 2MRS catalogue, as well as supplementary catalogues are presented in \S4. Discussions and conclusions can be found in \S5. {Throughout this paper we adopt $H_0=73$ km~s$^{-1}$ Mpc$^{-1}$, $\Omega_M=0.3$, and $\Omega_{\Lambda}=0.7$.}

\section{Group Finder}
\subsection{Friends-of-Friends}
The FoF algorithm has been the canonical method for identifying groups in magnitude-limited redshift-surveys because of its simplicity  \citep{Huchra1982,Ramella1997,Huchra2007,Duarte2014,Tully2015,Tempel2016}. Although the core algorithm remains, for the most part, unchanged from the original formalism by \cite{Huchra1982}, its subsequent applications to various redshifts surveys have varied according to their wavelength coverage, completeness, depth, and area. 

We based our FoF algorithm for the $K^{\rm o}_{\rm s}<$11\fm75 2MRS sample on the one developed by \cite{Huchra2007}, who created a group catalogue based on the 2MRS  $K^{\rm o}_{\rm s}<11\fm 25$ sample \citep{Huchra2005}. In its simplest form, this algorithm iteratively isolates galaxies $i$ and $j$ and determines whether they are in close enough proximity to each other in projected spatial and radial velocity-space to be gravitationally associated. If $i$ and $j$ are close in redshift space they are considered ``friends''. All friends of $i$ can be found this way, followed by all the friends of friends of $i$, and so forth. A group is then defined according to degrees of association. 

In principle we want galaxies to be considered ``friends'' if they are gravitationally bound. If the projected separation of galaxy $i$ and galaxy $j$, with an angular separation $\theta_{ij}$, defined as $D_{ij}=\sin\left(\frac{\theta_{ij}}{2}\right)\frac{v_{\text{avg}}}{H_0}$, is less than some linking length $D_l$, we know that they are close enough, in terms of a plane-of-sky projection, to be associated. 

We also consider the line-of-sight distance, represented in this case by the difference in $cz$ between the two galaxies ($\Delta v=|v_i-v_j|$). If this value is smaller than some linking velocity $v_l$, the galaxies are close enough to be associated along the line-of-sight. 

If a pair of galaxies meets both conditions (associated both along the plane-of-sky and line-of-sight) we consider them ``friends''. The two parameters used to determine whether two galaxies are friends are $D_l$ and $v_l$.

The simplest approach would be to keep these values constant. However, this does not take into account a variety of biases and only corrects for the change in projected separation with distance \citep{Huchra1982}. Following their work, as well as \cite{Huchra2007}, we have instead scaled $D_l$ in a manner which takes the sampling density as a function of redshift into account, given the survey magnitude limit. Namely, 
\begin{equation}
D_l=D_0\left[\frac{\int\limits_{-\infty}^{M_{ij}}\Phi(M)dM}{\int\limits_{-\infty}^{M_{\text{lim}}}\Phi(M)dM}\right]^{-\frac{1}{3}},
\label{eq:Dlink}
\end{equation}
where
\begin{align}
M_{\text{lim}}&=m_{\text{lim}}-25-5\log\left({v_f}/{H_0}\right),\\
M_{ij}&=m_{\text{lim}}-25-5\log\left({v_{ij}}/{H_0}\right), 
\end{align}

\noindent and $H_0$ and $m_{\text{lim}}$ are the Hubble constant and the apparent-magnitude limit of the sample, respectively, while $\Phi(M)$ is the \citet{Schechter1976} luminosity function, defined as:
\begin{equation}
\Phi(M)=C\ \Phi^*\ 10^{0.4(\alpha +1)(M^*-M)}\exp(-10^{0.4(M^*-M)}).
\end{equation}

\noindent where $C=0.4\ln(10)$. As proposed by \cite{Kochanek2001}, we adopt $\alpha=-1.02$, $M^*=-24\fm 2$, $\Phi^*=0.42\times 10^{-2} \text{ Mpc}^{-3}$ as the parameters of the $K$-band luminosity function for $H_0 = 73 $ km s$^{-1}$ Mpc$^{-1}$.

The fiducial velocity $v_f$ is set to $10^3$ km~s$^{-1}$, following \citet{Huchra2007}. This value was carefully determined in \cite{Huchra1982} by evaluating the probable number of interlopers as a result of the chosen value using the methods presented in \cite{Davis1982}. This particular value has been thoroughly verified by \cite{Geller1983,Ramella1989,Ramella1997}.

Defining the linking-length $D_l$ according to Eq. \ref{eq:Dlink} scales the surveyed volume by the number density of galaxies which can maximally be observed at that distance \citep{Huchra1982}. 
In the case of $v_l$, a scaling similar to that of Eq. \ref{eq:Dlink} could be implemented; however, it has been argued that this is unnecessary because the velocity dispersion of a galaxy group is independent of redshift \citep{Huchra1982}. Thus, setting $v_l$ to some constant value is sufficient and will not introduce a bias in velocity dispersion as a function of distance \citep{Huchra1982,Huchra2007}. Therefore, we set $v_l=v_0$, where $v_0$ is a constant.  

$D_0$ can be represented as the radius of the sphere used in calculating the density contrast ($\delta \rho / \rho$) in Eq. \ref{eq:D0} \citep{Huchra1982}: 
\begin{equation}
    \frac{\delta \rho}{\rho}= \frac{3}{4 \pi D_0^{3}}\left[\int\limits_{-\infty}^{M_{\text{lim}}}\Phi \left(M\right)dM\right]^{-1} -1 
    \label{eq:D0}
\end{equation}

Determining the appropriate values of $D_0$ and $v_0$ requires careful selection and optimization. If $D_0$ and $v_0$ are too small, all the galaxies will be put into their own individual groups. Similarly, in the extreme case, if $D_0$ and $v_0$ are too large, every galaxy will be put into one giant group consisting of the whole data set. The optimal choice of these parameters lies somewhere in between. A physical case can be made for mostly all choices, but a certain level of arbitrariness cannot be avoided. While the shortcomings (described below) of adopting fixed parameters have been accepted by previous works, we propose a new method {of} spanning through the FoF parameter space and using it to statistically identify groups. 

\subsubsection{Shortcomings of the traditional FoF algorithm}
Selecting a particular set of values for $v_0$ and $D_0$ can statistically alter the results when applying the traditional FoF algorithm. 

In some cases it may be appropriate to use tighter parameters when groups are small. However, larger groups may then be ``shredded'' into smaller, non-physical groups. Likewise, with a larger set of parameters, two physically distinct groups may be found as belonging to one group. A static parameter choice is unable to deal with the range in size, density, and dispersion of galaxy groups and clusters. 
Runaway may also occur in systems where galaxies are incorrectly included within groups because of their chance alignments between neighboring groups and other large-scale structures. This results in groups which are too large and often too elongated to be considered physical.   

Furthermore, the FoF algorithm renders all results (groups) with equal confidence. Groups with a large number of members which are very tightly constrained in redshift space are considered equally likely than groups consisting of a very low number of members that are loosely constrained in redshift space.

For these reasons, we have developed a modified version of the FoF that considers many different sets of linking lengths, thereby rendering the group finder statistically robust. 

\subsubsection{Hard limits}

To prevent runaway connections between groups, we introduce hard limits in both redshift space $(v_{\text{max}})$ and projected on-sky distance $(D_{\text{max}}$). This constrains how large any one particular group can grow. After the first friends are found, the center of the proto-group $(\bar{\alpha},\bar{\delta},\bar{v)}$ is determined, where these parameters are the mean values of RA, Dec, and velocity of the current group members, respectively. Friends-of-friends are only added for galaxies within the proto-group with $|v_i-\bar{v}| < v_{\text{max}}$ and with projected angular distance $\theta_i = \sin\left(\frac{\Delta\theta_i}{2}\right)\frac{v_i}{H_0} < D_{\text{max}}$, where $\Delta \theta_i$ is the angular separation between the galaxy $i$ and the center of the proto-group. The center of the proto-group is recalculated with every new addition of a member until no new members are found. 

Although $v_{\text{max}}$ and $D_{\text{max}}$ are not dissimilar to the linking lengths $v_l$ and $D_l$, their choice is not as physically ambiguous. The values $v_{\text{max}}$ and $D_{\text{max}}$ can be thought of as the maximum velocity distribution of a group and the maximum radius of a group, respectively. Both properties have been studied extensively for well-known groups and clusters \citep{Dressler1980,Huchra1982,2df2001,6df2004,Huchra2007,Robotham2011,Mehmet2014,Ramatsoku2016}. We set $D_{\text{max}}$ to the Abell radius, $R_A = $ 1.5 $h^{-1} = 2$ Mpc \citep{Abell1989}. Typical group and cluster velocity distributions ($\sigma_v$) have been shown to vary between $500$ km~s$^{-1}$ and $1000$ km~s$^{-1}$ \citep{Peebles1980,ShapleyBook1981,Dressler1988,Abell1989,Ramella1997,loeb2008,Tempel2016,Coutinho2016}. We therefore set $v_{\text{max}}$ to represent the 3$\sigma$ level for a typical cluster, i.e. 3$\sigma_v$= 3000 km~s$^{-1}$. The limit along the line-of-sight direction is purposely large in order to account for the observational bias associated with redshift measurements in that direction. The higher quality accuracy in the plane-of-sky positions of the galaxies allows for a stricter limit. 

The hard limits are chosen to be the size of distinct clusters since this is the largest that a group could physically grow. This ensures that ``shredding'' of clusters and particularly large groups does not occur, while at the same time inhibiting runaways. Smaller groups will still be identified.  

By introducing these hard limits, the modified FoF algorithm will migrate towards the highest densities of galaxies first rather than around randomly-chosen galaxies. This has the advantage that large collections of galaxies will preferentially shift the center to the cores of groups, reducing any bias that the algorithm may have with respect to the randomly-selected starting galaxy.

In summary, the introduction of hard limits improves the FoF algorithm by inhibiting runaways, allowing large groups and clusters to be found without shredding them into smaller groups, and by increasing its sensitivity to higher densities rather than randomly-selected starting points.

{An unfortunate drawback from introducing hard limits is that the group finder is not robust to initial conditions. The order which the group finder is run alters the results. To correct this (as well as some other issues), the group finder is stabilized over many runs.}

\subsubsection{Stabilizing the algorithm over many runs}
To stabilize the algorithm we execute many different runs of the modified FoF algorithm (with hard limits) for varying sets of linking length parameters and random starting points. The results of the numerous runs are averaged and conglomerated into one final result. Accepting groups which are found over several runs ensures that the results are robust to initial conditions and that statistical anomalies are removed. Secondly, by working with a range of parameters, larger groups and massive clusters that are not a chance agglomeration of smaller groups can still be identified without losing sensitivity to smaller groups. The execution of numerous runs furthermore allows the construction of confidence intervals. Associations of galaxies found under the strictest of parameters are more reliable than ones based on a more relaxed choice of parameters.   

Every run produces a new group catalogue but not every group appears in every run. Groups can be split up, combined, have varying members, or disappear completely from run to run. Because of this, we focus on tracking galaxy-galaxy pairs throughout all the runs. Since the defined galaxy groups themselves change with every run, they cannot be as easily tracked. However, the number of times galaxy $i$ and galaxy $j$ are found in a group together can be tracked reliably through any number of trials. In this way we are able to assess the strength of the connection between any galaxy pair.  

Recording galaxy-galaxy associations results in a list of vectors. A single vector comprises the $i^{th}$ and $j^{th}$ galaxies, and the integer parameter $w$ describes how many times galaxies $i$ and $j$ were found to be members of the same group. After $k$ runs on a data set containing $n$ galaxies, a final list of  $n(n-1)/2$ elements is created where each element takes the form of $(i,j,w)$ where $i,j \in [1,n], i \neq j $ and $w \in [0,k]$. This list can then be used to recreate the groups, taking into account the varying parameter space covered throughout the runs. This type of data set, being topological in nature, lends itself well to graph-theory which we use to interpret pairwise interactions.

\subsection{FoF Graph Theory adaptation}

\begin{figure}
    \centering
    \includegraphics[width=0.5\textwidth]{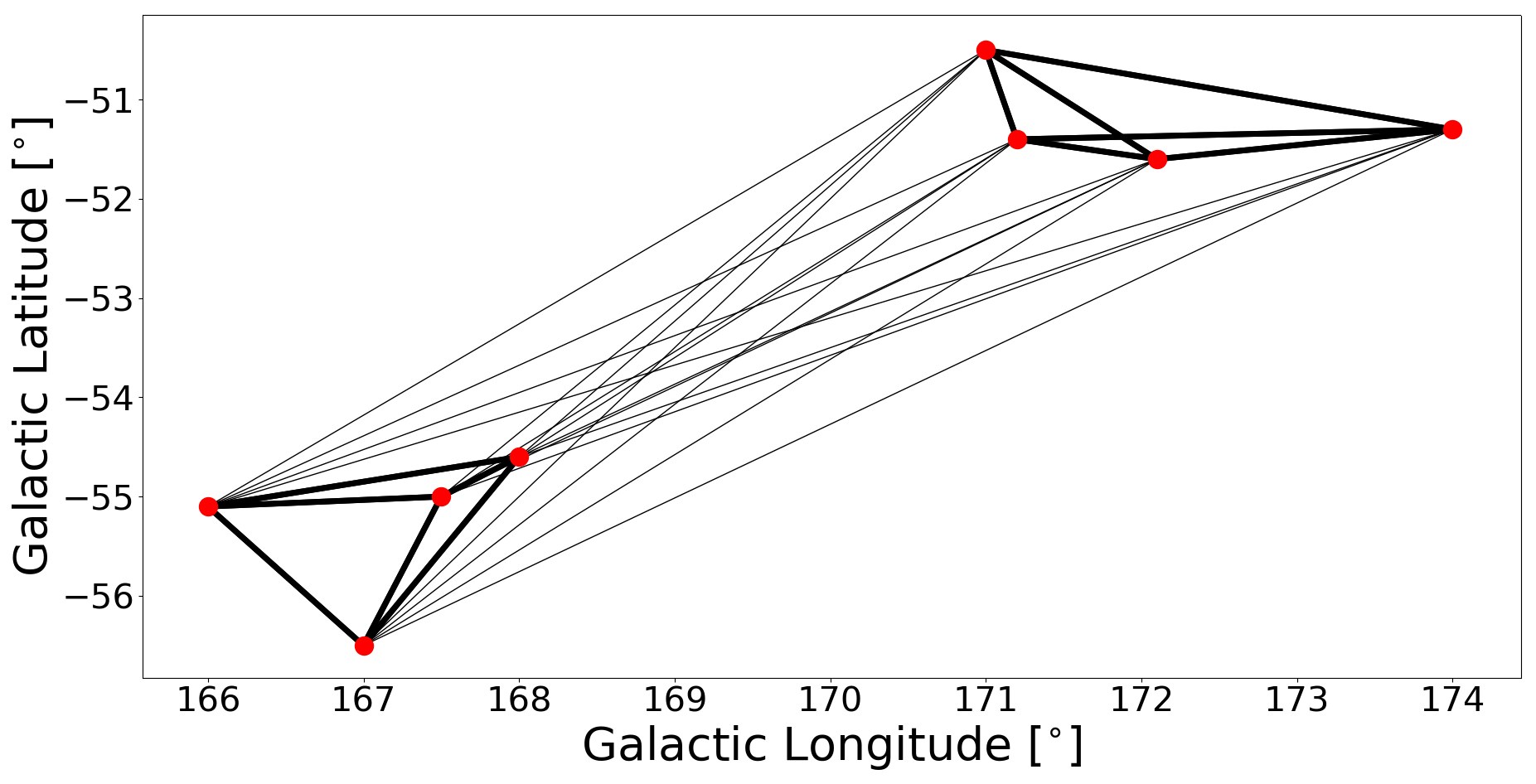}
    \includegraphics[width=0.5\textwidth]{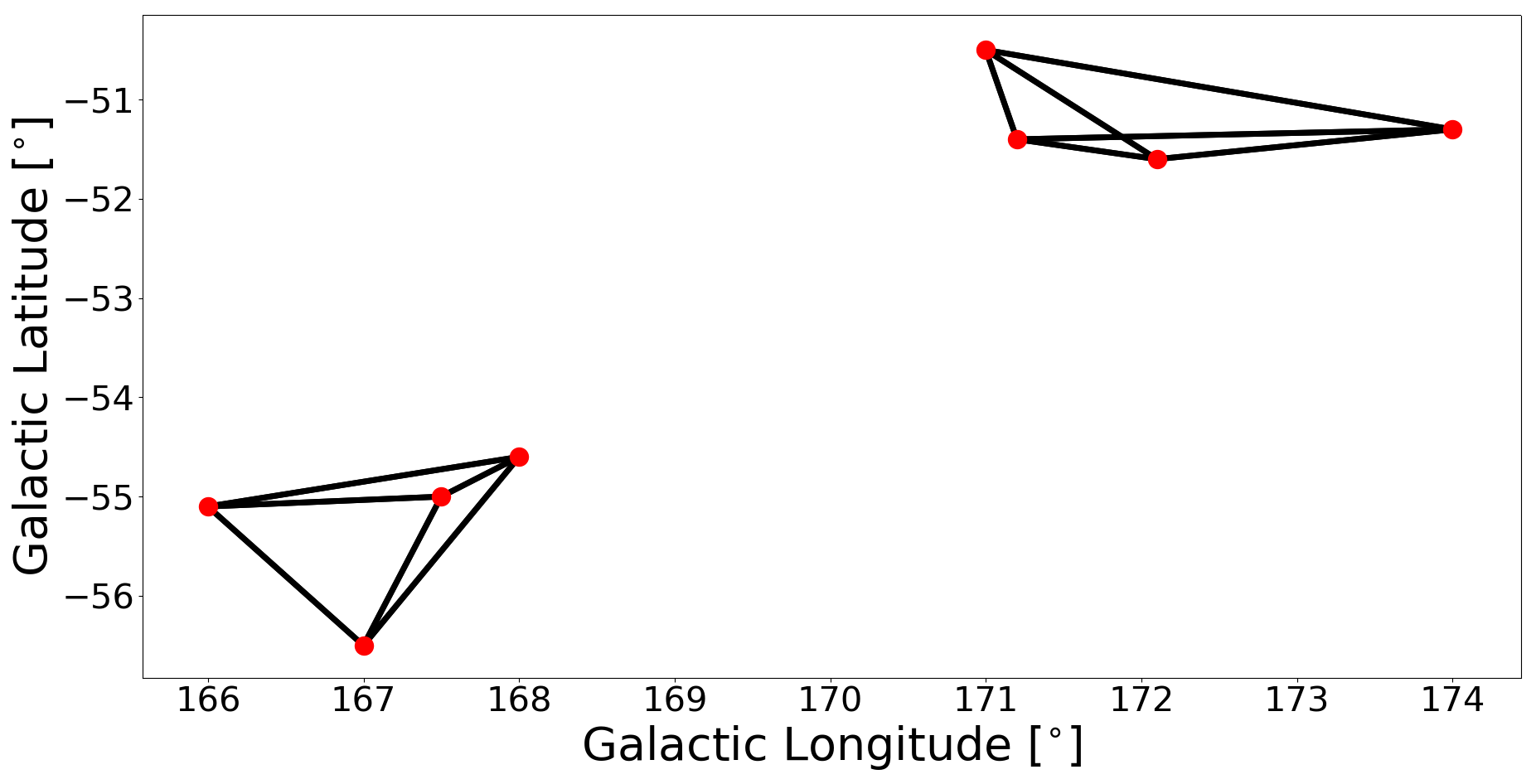}
    \caption{Graph representation of a single group after averaging over many runs. Points represent galaxies in the 2MRS survey. Edges are weighted according how often the connected pair were found in the same group. Thick edges represent pairs where galaxies were found $> 90 \%$ of the time, thin edges pairs which were identified $< 10\%$. The top panel shows the group before weak edges are removed. The bottom panel shows how the group is split into two groups after the weak edges were removed. }
    \label{fig:graph2}
\end{figure}

To manage and assess the resulting group catalogues that we average over, a mathematical graph is constructed from the final list of FoF results. The nodes of the graph represent the galaxies in the redshift survey, and the connecting lines, or weighted edges between the nodes, represent the number of times a pair of galaxies is found in the same group. Due to the topological nature of the resulting data set, mathematical graphs are an optimal way to manage and quantify results from multiple runs, and provide a powerful visualization tool \citep{Pascucci2011Book}. To ensure stability against statistical anomalies, edges with a $w$ value less than 0.5 (i.e., 50\% repeatability) are removed. This results in a single large graph which consists of many smaller disconnected subgraphs. The final group catalogue is generated by identifying all the disconnected subgraphs, indicated by isolated objects within the main graph, using the \texttt{Graph.subgraph} function in the \texttt{networkx} Python package. The remaining subgraphs are the groups selected for further inspection. At this point every group in the catalogue can be represented as an independent, self-contained graph. 

An example of how removing weak edges results in statistical robustness is shown in Fig.~\ref{fig:graph2}, where the graph represents an average over many runs for a particular group. The points (galaxies) are plotted in Galactic coordinates. Visual inspection of the points imply two separated groups; note that they were found as a single group in less than $10\%$ of the runs. This can be seen from the weighting of the weakest edges (lines) connecting the two groups. Edges are also closely related to the linking length choice; in this example, the algorithm finds one group for a large linking length while a more moderate linking length results in two separate groups. We run a continuous range of linking lengths ($D_0=0.56 \pm 0.1$~Mpc and $v_0=350 \pm 100$~km~s$^{-1}$) over several runs. This allows the removal of weak edges, hence eliminating statistical anomalies and rejecting non-physical linking lengths. 

This visualization technique provides a synergistic medium between human and computer understanding. A simple inspection of the the points in Fig.~\ref{fig:graph2} (in this particular example) clearly demonstrates that these two groups really are separate entities when overlaid with weighted edges. Averaging over many runs using graph theory and only considering statistically significant groupings and physical linking lengths (by removing weak edges) results in a stable and robust algorithm. In other words, the same group catalogue is generated regardless of the initial starting point of the algorithm or the order in which the algorithm cycles through the galaxies -- an outcome which is not guaranteed in the traditional FoF algorithm.

Edges are powerful tools; however, a lot of additional information can also be extracted from individual nodes in the respective group graphs which provide another angle of visual interrogation. In a similar way to edges, the nodes of all the graphs are weighted to take into account their number of connections and how strong said connections are. This is accomplished by defining the ``connectedness score'' of an individual node in a graph as:
\begin{equation}
    S_{i} = C_i W_i, \\
\label{eq:Score}
\end{equation}
\noindent where $C_i$ is the percentage of the graphs edges connected to node $i$ and $W_i$ is the average weight of all the edges connected to node $i$, defined respectively as
\begin{equation}
    C_i = c_i\left[\frac{n\left(n-1\right)}{2}-1\right]^{-1}
\end{equation}
\noindent and 
\begin{equation}
    W_i= {\sum \limits_{j=1}^{c_i} w_{ij}}/{c_i},
\end{equation}
where $c_i$ is the total number of edges connected to node $i$, $n$ is the total number of nodes in the graph, and $w_{ij}$ is the weight of edge $j$ connected to node $i$ (the percentage of runs in which galaxy $i$ was found in the same galaxy group as galaxy $j$. 

We define two metrics with respect to the graph itself, namely the total and average connectedness score. These two global group statistics quantify the reliability of a group given the strength of its edges and the number of members. Indeed, the total connectedness score can be thought of as the corresponding number of nodes which make a complete graph (a graph where each node is connected to every other node) while the average connectedness score would represent the number of members of a group found in every single run. For example: a group might have 20 group members, with a total connectedness score of only 10. This would imply that it could be a group of 10 members found in every single run, with an additional 10 members found at a lower weighting or connectedness, while still satisfying the criterion of at least 50$\%$ repeatability. 

These are the metrics that we later use to identify groups that seem unstable and require further inspection by visual examination. 

\subsection{Identifying Substructure}
\label{seq:SubStructures}
The graph description of each individual group furthermore allows for an innovative method of identifying substructure within groups, therefore extending the capabilities of the original FoF algorithm. 

Cutting at a fixed significance level ensures robustness. However, cutting at higher significance levels can, in some cases, result in a group being sub-divided further. In order to combat the pseudo-randomness of the significance cut, we inspected every group to determine whether a stricter significance cut would result in separating groups. If this is the case, we identify these as substructures in a given group. 

For every group, edges are iteratively removed from the graph from lowest to highest weighting, and with every iterative cut the number of subgraphs with three or more nodes is calculated. Once all edges have been removed, the maximum number of subgraphs with the lowest edge-cut become the subgroups. 

\begin{figure}
    \centering
    \includegraphics[width=0.5\textwidth]{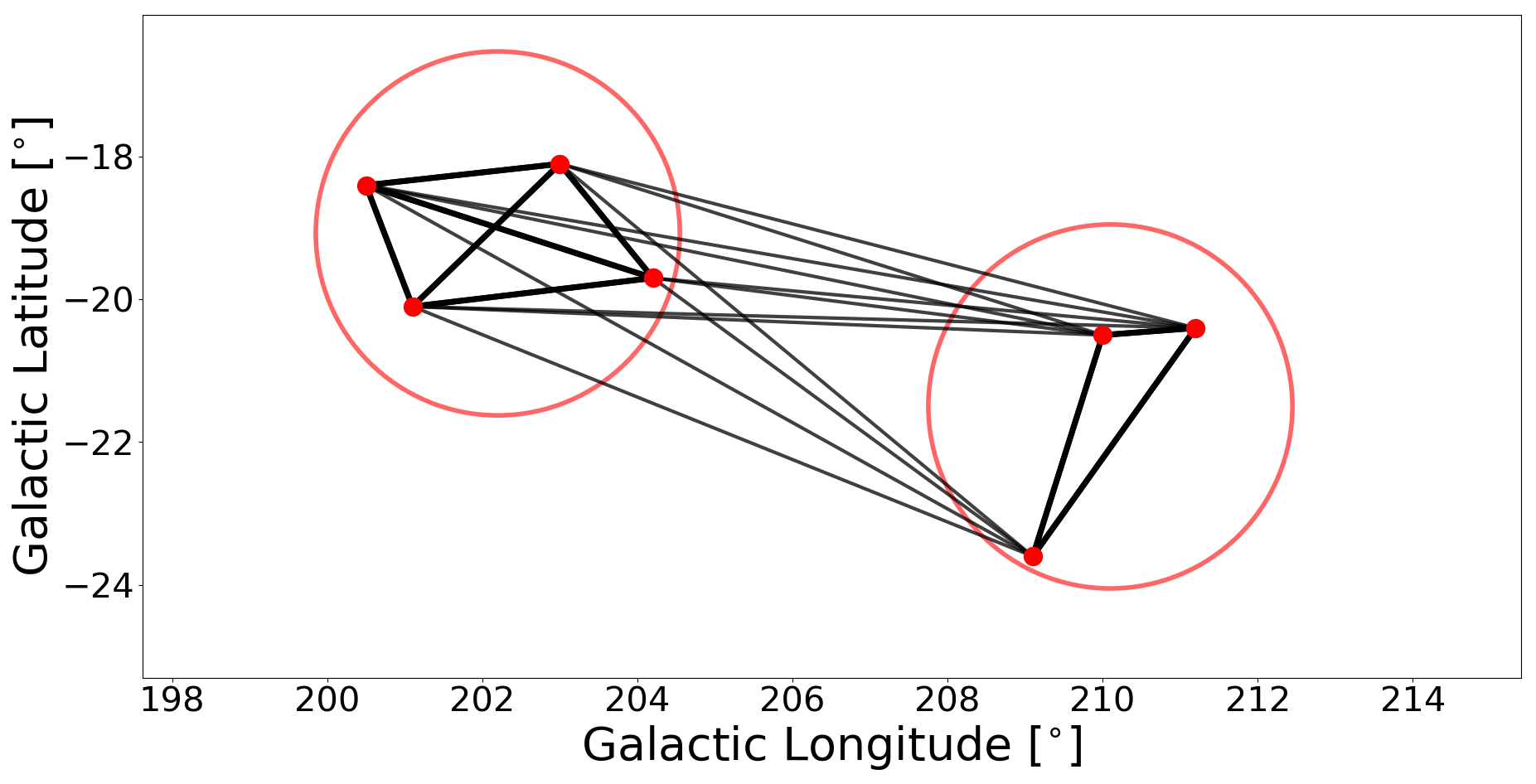}
    \caption{Graph representation of a group with sub-groups. Nodes represent galaxies in the 2MRS. Edges are weighted according to the percentage of the runs in which a galaxy pair is found for the same group. Strong edges represent $>90 \%$, weak edges between $60 \% - 70 \%$. The two circles show the two subgroups identified when using our method.}
    \label{fig:subgroup_example}
\end{figure}

Fig.~\ref{fig:subgroup_example} shows an example of such a case. The entire system is identified as a single group because all edges in Fig.~\ref{fig:subgroup_example} are within the cut-off limit of 50\%. However, if the cut-off limit had been set to 70\%, then the weak edges would have been removed, resulting in two independent groups (in the same way as the groups in Fig.~\ref{fig:graph2}). Removing edges sequentially from weakest to strongest results in two separate structures being identified. These substructures are classified as subgroups. In Fig.~\ref{fig:subgroup_example} the subgroups are represented by the two large red circles.

\subsection{Correcting for Unreliable Connections}
A lot of effort has been put into addressing shortcomings of the FoF algorithm, since it is so easy to connect nonphysical groups because of chance alignments. The graph method provides a powerful tool to correct for this. Nevertheless, a remnant of this issue remains for more complicated systems (such as the Virgo Supercluster); it is still possible for two or more obviously disconnected groups to be assigned to a single group and in some cases a low number of edges/connections can remain after a significance cut. 

In some cases, visual evidence seems to suggest that two groups may be distinct and are arbitrarily connected through a single connection. However, computationally the entire system is seen as a single group and it is non-trivial to separate it into two distinct entities. This problem is compounded by the fact that there might be more than one unreliable connection. Determining what is an acceptable level of connections between these two groups before they are thought of as a single group is a challenging exercise. Furthermore, removing single edges and looking for a split in the group quickly becomes computationally exhaustive and is near impossible for cases where there is more than one connection. 

In order to solve this issue we looked at the average connectedness score of every group and whether or not those groups had any subgroups. The average connectedness of groups becomes a useful metric to flag these cases: when two obviously-separated group have very few connections, the number of connections in the group are less than the total number of possible connections. Thus, the average connectedness score is very sensitive to such situations and looking for groups with low values of this statistic that have subgroups immediately allows us to flag this issue. 

While this artifact exists for less than 1$\%$ of the total groups, it is still important to correct for it. The best solution is to simply allow the subgroups to become their own groups. Any galaxies not identified as belonging to a subgroup were assigned to the nearest on-sky subgroup, as long as the value of $R_{200}$ (defined in Eq.~\ref{eq:R200}) of that subgroup was greater than the distance to the left-over galaxy.

\begin{figure}
    \centering
    \includegraphics[width=0.49\textwidth]{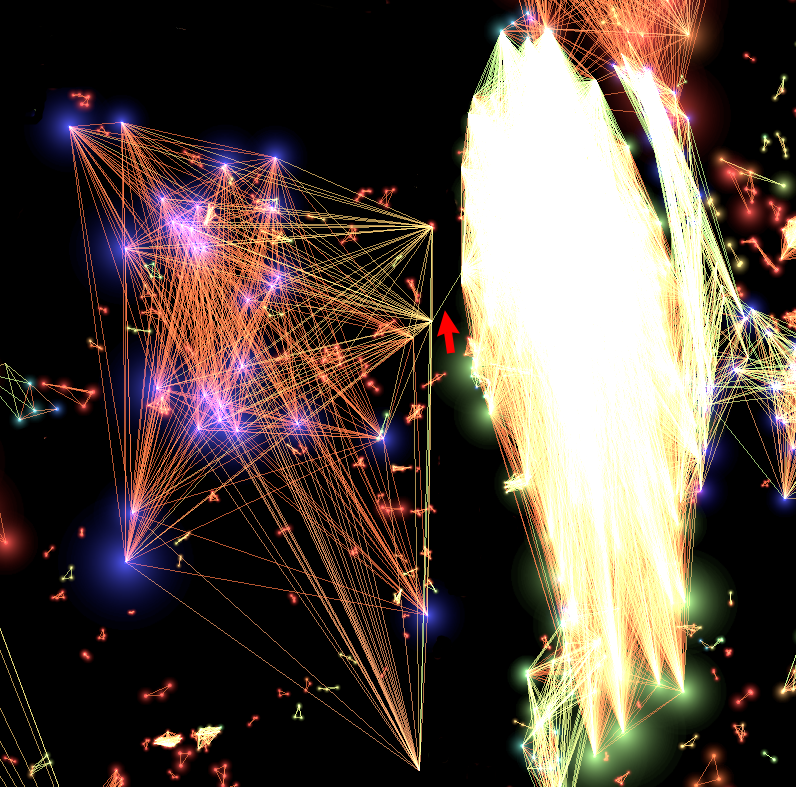}
    \includegraphics[width=0.49\textwidth]{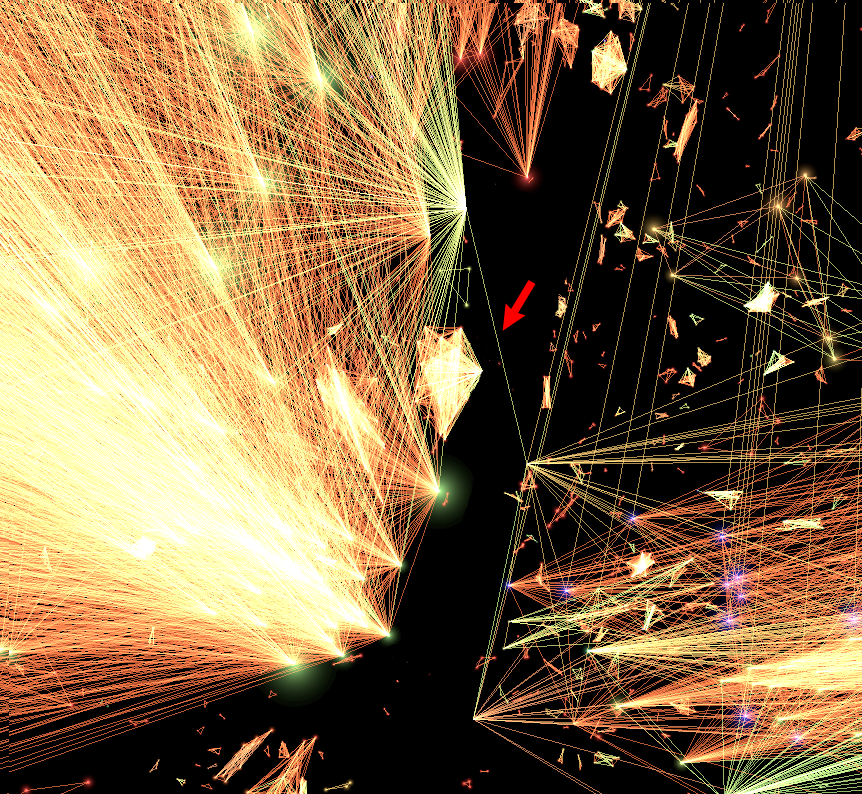}
    \caption{3D visualization of the Virgo Supercluster. The two large (independent structures) are both identified as a single group before corrections (as described in Sect.~2.4 are made. Green, orange, and red points represent galaxies with a scores of $>70\%$, $>80\%$, and $>90\%$, respectively. The top and bottom panels shows a normal and zoomed-in view of the false connection, respectively.}
    \label{fig:Bad_Virgo}
\end{figure}

Figure~\ref{fig:Bad_Virgo} shows the 3D representation of the Virgo Supercluster as found in the 2MRS. There are two visually evident structures, which were identified as a single group before applying the correction described in \S2.4. The edges shown in Fig.~\ref{fig:Bad_Virgo} are all $>70\%$ and both structures are connected to one another because of a single connection occurring at the $>70\%$ level. Thus, applying increasingly stricter cuts does not split the independent structures. This is the problem that the correction described in \S2.4 aims to address. After this correction all three structures are identified as their own independent groups. Besides the main Virgo cluster, the now independent groups were identified as Virgo \MakeUppercase{\romannumeral 2} groups (in particular the M61 and NGC4753 groups at $l=298.99^{\circ}$ and $b=66.3 ^{\circ}$ and the NGC4697 and NGC4699 groups at $l = 306.11^{\circ}$ and $b=53.7^{\circ}$) and form part of the Virgo southern extension~\citep{Nolthenius,Giuricin2000,Karachentsev2013,Kim2016}.  
\subsection{Linking Lengths}

{\cite{Ramella1989} measured the number of groups, total number of galaxies in groups, the median velocity dispersion, and the first and third quartiles of the median velocity dispersion for the $6^{\circ}$ and $12^{\circ}$ CfA \citep{Huchra1982} survey as well as a mock catalog of the survey by \cite{deLapparent1986}. This was done for varying linking lengths, with $D_0=0.56$ Mpc and $v_0=350$ km~s$^{-1}$ found to be the optimum choice of linking lengths at limiting the number of interlopers but not breaking apart large clusters. These same optimum linking lengths were later verified again in \cite{Ramella1997} for the Northern CfA2 survey \citep{Huchra1995}. They measured numerous group properties for several different linking lengths, which were compared to a geometric simulation of the sample. The linking lengths were also validated in \cite{Ramella2002} on both the Updated Zwicky Catalog and Southern Sky Redshift Survey \citep{Falco1999,Ochesenbein2000}. Later, this combination of linking lengths was applied successfully to earlier versions of the 2MRS \citep[namely][]{Huchra2007}. In summary, this linking length choice of $D_0=0.56$ Mpc and $v_0=350$ km~s$^{-1}$ has been successfully used over four different surveys (including an earlier 2MRS survey) and has been validated by several independent methods.}

{The parameter choice is discussed further in the next section by evaluating it against our own mock catalog.} In order to incorporate the graph structure, we vary $D_0$ and $v_0$ between $0.56 \pm 0.1$~Mpc and $350 \pm 100$~km~s$^{-1}$, respectively, over 100 runs using 2 kpc and 2~km~s$^{-1}$ steps, respectively). The larger linking lengths will accommodate large clusters with dispersions of 1000 km~s$^{-1}$ given their larger gravitational potential wells.    

Our choice of linking length parameter was validated against a 2MRS-like mock catalogue, generated by the Theoretical Astrophysical Observatory \citep[TAO,][]{Bernyk2016} described in the next section. The mock is complete up to $K_{\rm s}^{\rm o} = 14\fm75$, i.e. three magnitudes deeper than the 2MRS and contains both cosmological and peculiar redshifts.

\subsection{Validation of Algorithm on a Mock Catalogue}

{The TAO mock catalog was used to (1)  evaluate the choice of linking lengths set by \cite{Ramella1997} and (2) substantiate how accurately groups are recovered in the 2MRS given the relatively modest magnitude limit of $K_{\rm s}^{\rm o}<11\fm 75$ and the effect of incompleteness as a function of magnitude limit and observed recession velocity.}

\subsubsection{Validating Parameter Choice}
{The group finder was run with numerous different linking length pairs and compared to groups which were identified in the cosmological TAO mock catalog (which are free from redshift distortions). Two standard metrics used in optimizing group finding algorithms are ``completeness" and ``reliability". Completeness is a measure of the percentage of known group members (from the mock data) which are recovered when running the algorithm. Reliability is a measure of the percentage of members found via the algorithm which actually belong to the same group. Ideally, an algorithm would be designed such that both completeness and reliability are maximized. However, optimizing an algorithm over two variables is non trivial. In most cases completeness and reliability are optimized via a single variable which is made up of simple combinations of reliability and completeness (such as the sum or product) \citep{Stothert2019}. Completeness and Reliability can be calculated as}
\begin{equation}
C=\frac{1}{\sum_{i=1}^{N_G} \sum_{j=1}^{N_H} n_{ij}}\sum_{j=1}^{N_H}\text{max}_i \left(n_{ij}\right)
\end{equation}
and 
\begin{equation}
R=\frac{1}{\sum_{i=1}^{N_G} \sum_{i=1}^{N_G} n_{ij}}\sum_{j=1}^{N_H}\text{max}_j \left(n_{ij}\right)
\end{equation}

\noindent{respectively, where $n_{ij}$ is the number of members found in both group $i$ (from the redshift distorted mock catalog) and group $j$ (mock catalog with real distances) and $N_{G}$ and $N_{H}$ represent the total number of groups in the redshift-distorted mock and the real-distances mock, respectively. }

{A closely related statistic to reliability and completeness is that of Variation of Information (VI), which is a measure of how much information of a data-set can be inferred from another and vice versa. It provides a measure of how independent they are and effectively combines completeness and reliability into a single metric \citep[see][for further discussion]{Melia2005,Meilia2007}. This method has been used successfully by \cite{Stothert2019} who defined VI as }

\begin{align}
\begin{split}
    VI=&-\sum_{j=1}^{N_H} \left( \frac{\sum_{i=1}^{N_G} n_{ij}}{\sum_{i=1}^{N_G} \sum_{j=1}^{N_H} n_{ij}} \ln\left(\frac{\sum_{i=1}^{N_G} n_{ij}}{\sum_{i=1}^{N_G} \sum_{j=1}^{N_H} n_{ij}} \right) \right) \\
    &- \sum_{i=1}^{N_G}   \left( \frac{\sum_{j=1}^{N_H} n_{ij}}{\sum_{i=1}^{N_G} \sum_{j=1}^{N_H} n_{ij}} \ln\left(\frac{\sum_{j=1}^{N_H} n_{ij}}{\sum_{i=1}^{N_G} \sum_{j=1}^{N_H} n_{ij}} \right) \right)\\
    &- 2 \sum_{i=1}^{N_G} \sum_{j=1}^{N_H} \left( \frac{n_{ij}}{\sum_{i=1}^{N_G} \sum_{j=1}^{N_H} n_{ij}} \ln \left( \frac{ \left(n_{ij}\right)\left(\sum_{i=1}^{N_G} \sum_{j=1}^{N_H} n_{ij}\right) }{ \left(\sum_{j=1}^{N_H} n_{ij}\right) \left(\sum_{i=1}^{N_G} n_{ij}\right) } \right) \right)
\end{split}
\end{align}

\begin{figure*}
    \centerline{
    \includegraphics[width=1.2\textwidth]{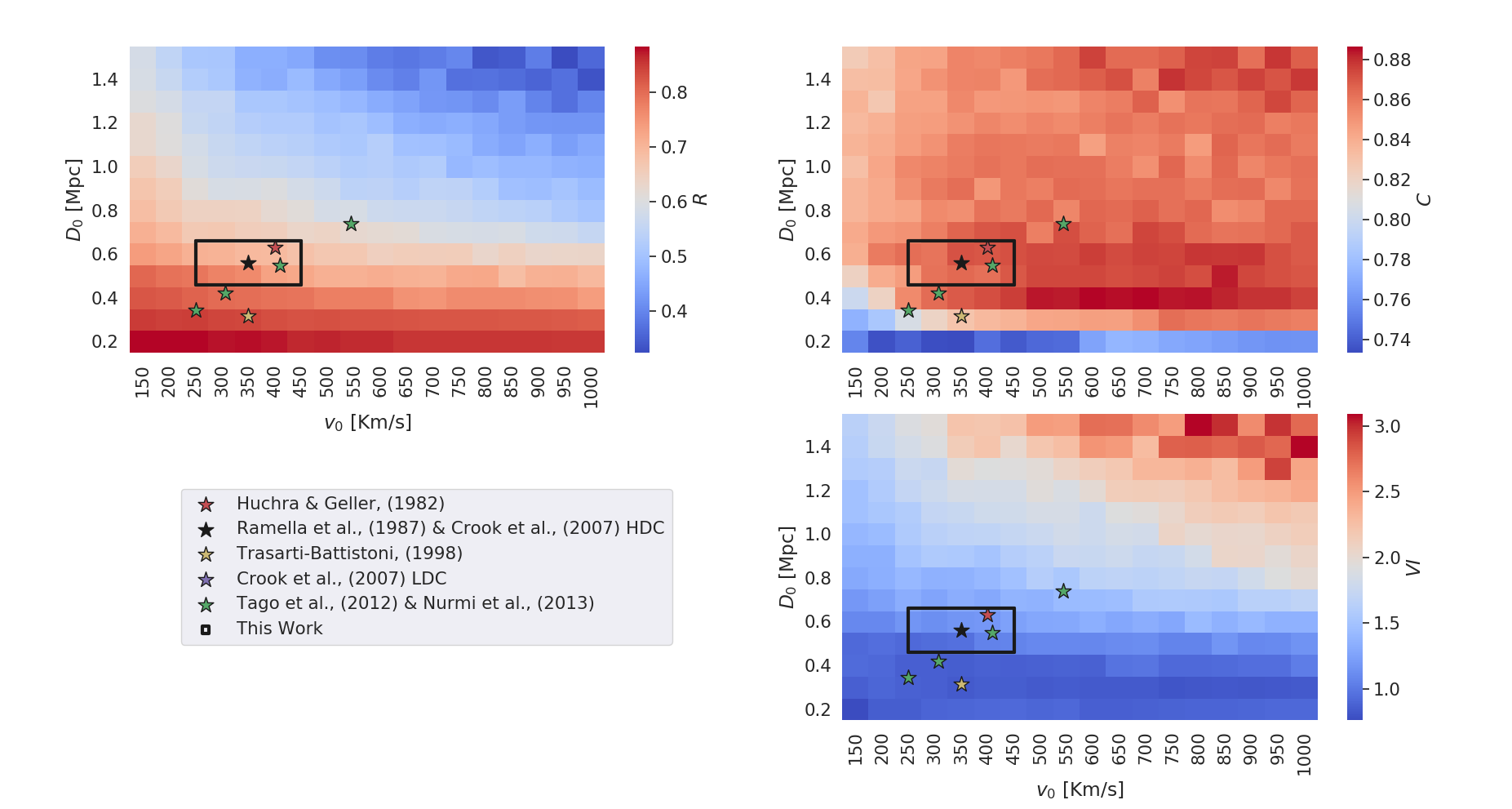}}
    \caption{{Completeness (top left), Reliability (top right) and Variation of Information (bottom right) results when comparing groups found in the purely-cosmological TAO mock catalog to groups found by running our group finder on the redshift-distorted TAO mock over several different parameter choices. Each pixel represents a comparison. The $v_0$ and $D_0$ parameters for each individual run form the x and y axes, respectively. Stars represent the parameters chosen by previous works. The black square shows the range of parameters probed by this work when implementing the graph theory method discussed in \S2.2.}}
    \label{fig:VI}
\end{figure*}

{All three metrics were used in order to justify our choice of central linking lengths, which coincide with the studies of \cite{Ramella1997} and \cite{Huchra2007}. The results of running different parameters, which are then compared to the groups in the cosmological mock, are shown in Fig. \ref{fig:VI}. The top-left and top-right panels show reliability and completeness, respectively. When linking lengths are strict the reliability score is large, whereas large linking lengths have the opposite effect. The inverse is true for the completeness metric: when linking lengths are small, fewer groups make the cut (reducing completeness); however, the groups that do make the cut are are recovered in the mock. Alternatively, when linking lengths are large, many groups pass the cut; however, the probability of false detections increases and a drop in reliability is seen. The completeness and the reliability results both provide a useful sanity check. Reliability tends to be lower than completeness in general. When visually inspecting the mock catalog, the effect of redshift distortions is evident. Chance alignment of galaxies, in combination with a large spread in velocity due to peculiar motions in groups and large-scale cosmic flows, cause several false detections. This effect cannot be removed and is a consequence of identifying groups in redshift space. It is important to keep this in mind when using this (or any other) redshift-based group catalog.}

{The $VI$ results show where we could optimize both completeness and reliability. We overlay the parameter choices used by several different authors also running FoF algorithms. Of particular interest is the black star, which represents the parameter choice used by \cite{Ramella1997} and later \cite{Huchra2007} that we adopt in this work. The range of parameters spanned by our group finder (making use of the graph theory adaptation discussed in \S2.2) is demarcated in Fig. \ref{fig:VI} as a black rectangle.}

{The \cite{Ramella1997} parameters result in a reliability $R>0.7$ and a completeness of $C>0.8$. These parameters also have low $VI$ scores, implying a reasonably good recovery of the mock groups. The \cite{Ramella1997} linking lengths are also close to values used by other authors. All linking lengths chosen by previous authors seem to hover around the $VI$ ``well" in Fig. \ref{fig:VI}.}

\subsubsection{Evaluating Incompleteness}

\begin{figure}
    \centering
    \includegraphics[width=0.5\textwidth]{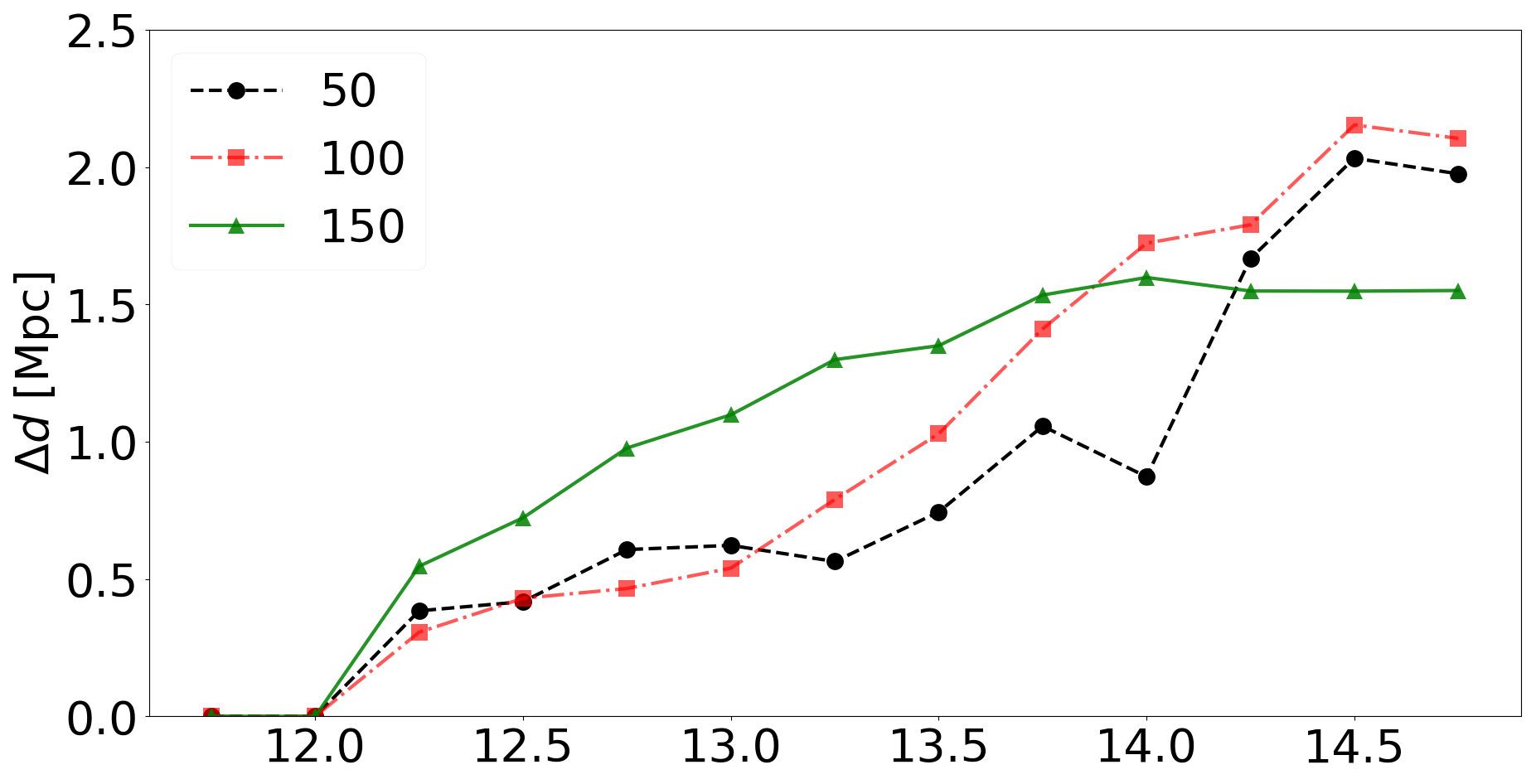}
    \includegraphics[width=0.5\textwidth]{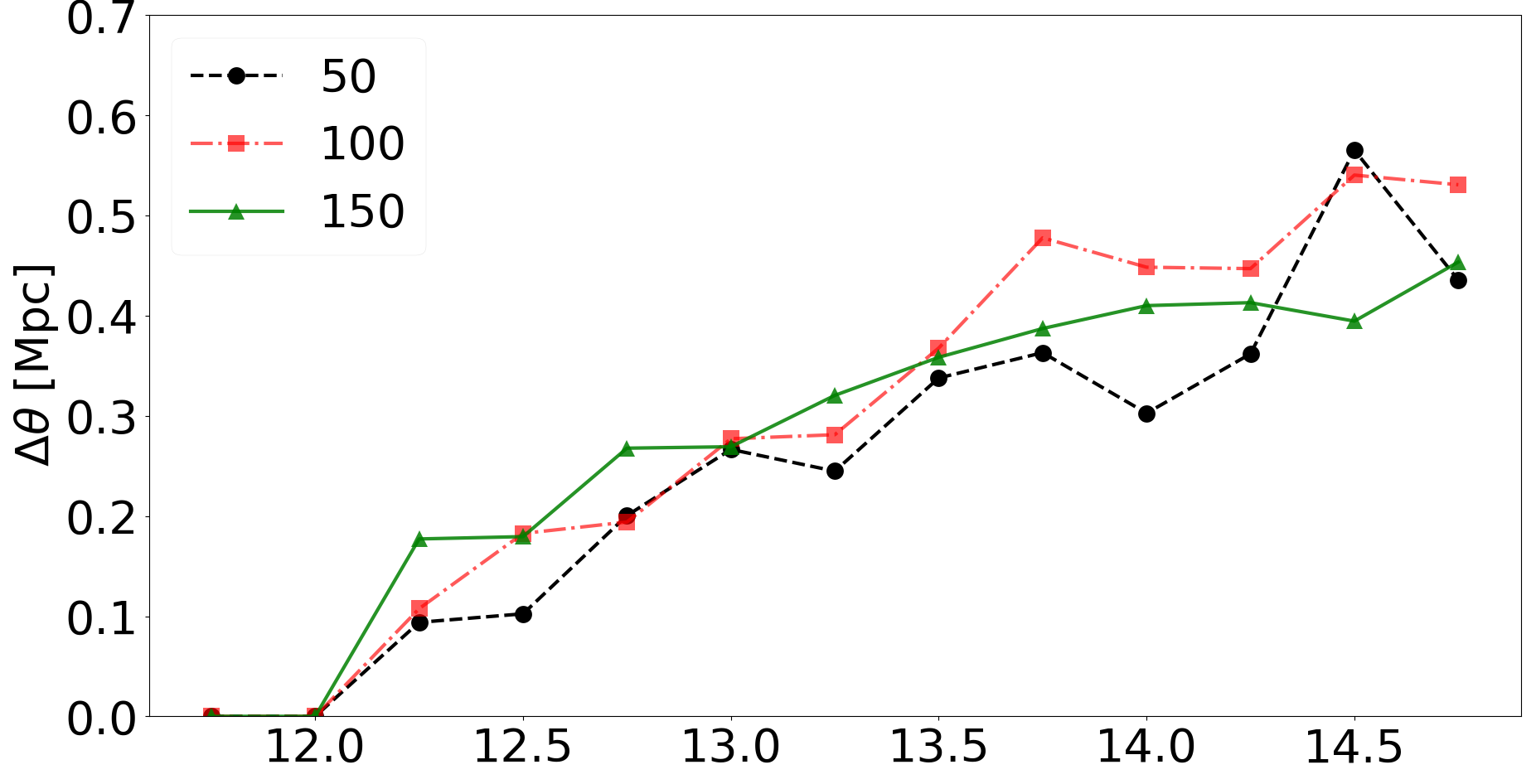}
    \includegraphics[width=0.5\textwidth]{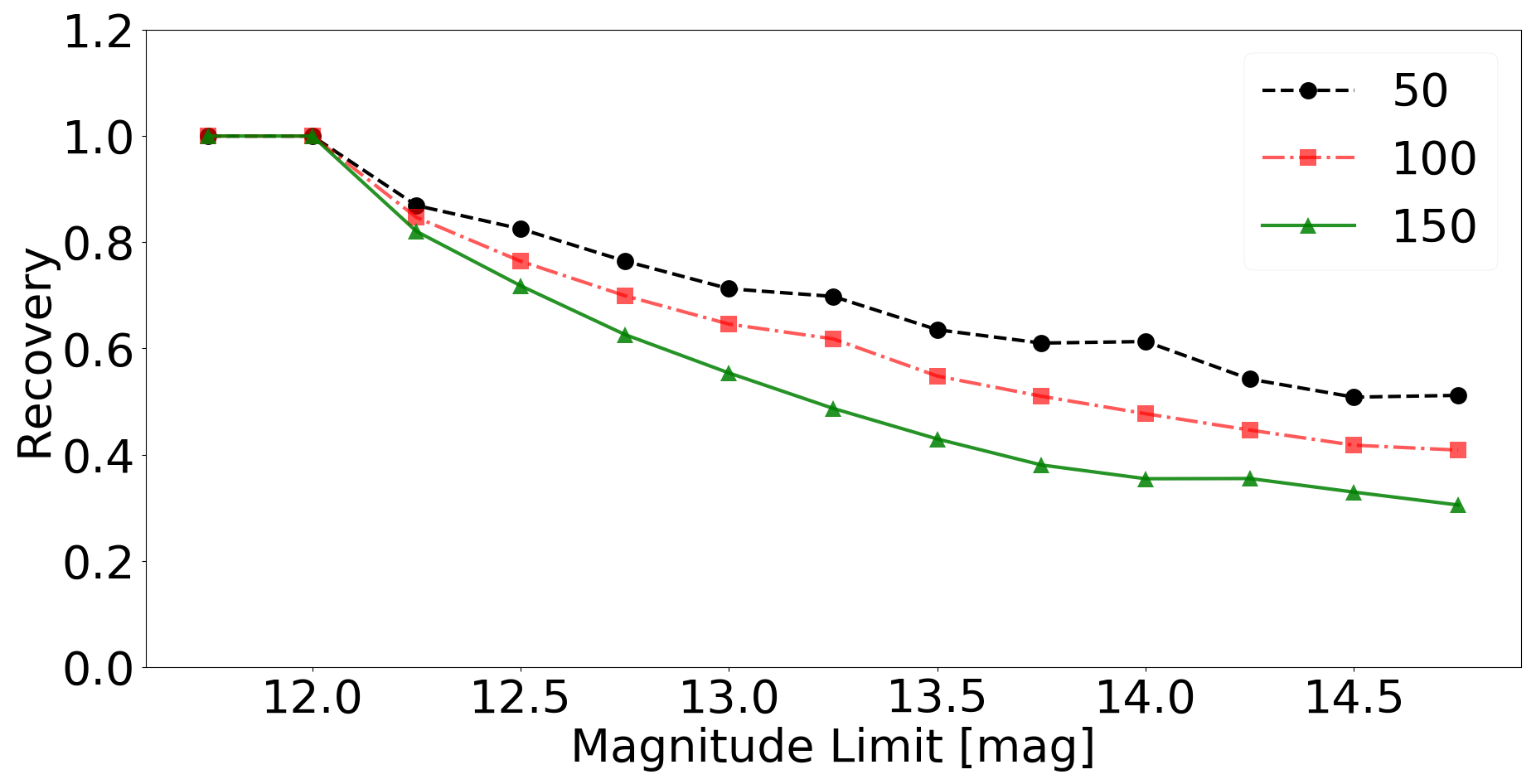}
    \caption{Trends of galaxy groups with three members found in the TAO mock catalogue at the 2MRS magnitude limit of $K_{\rm s}^{\rm o}<11 \fm 75$, as a function of magnitude completeness limit (x-axis) and distance (different curves). The solid, dashed and dotted lines represent groups with a co-moving distance of $0 < d \leq 50 $ Mpc, $50 \text{ Mpc}< d \leq 100$ Mpc, and $100 \text{ Mpc} < d < 150 \text{ Mpc}$, respectively. Top: offset in line-of-sight position with changing depth. Middle: projected on-sky offset with changing depth. Bottom: percentage of the total group recovered at a magnitude limit of $K<11 \fm 75$. This plot quantifies how accurately positions of groups are recovered and what percentage of the group members are found.}
    \label{fig:mock_plot}
\end{figure}

\begin{figure*}
    \centering
    \includegraphics[width=0.49\textwidth]{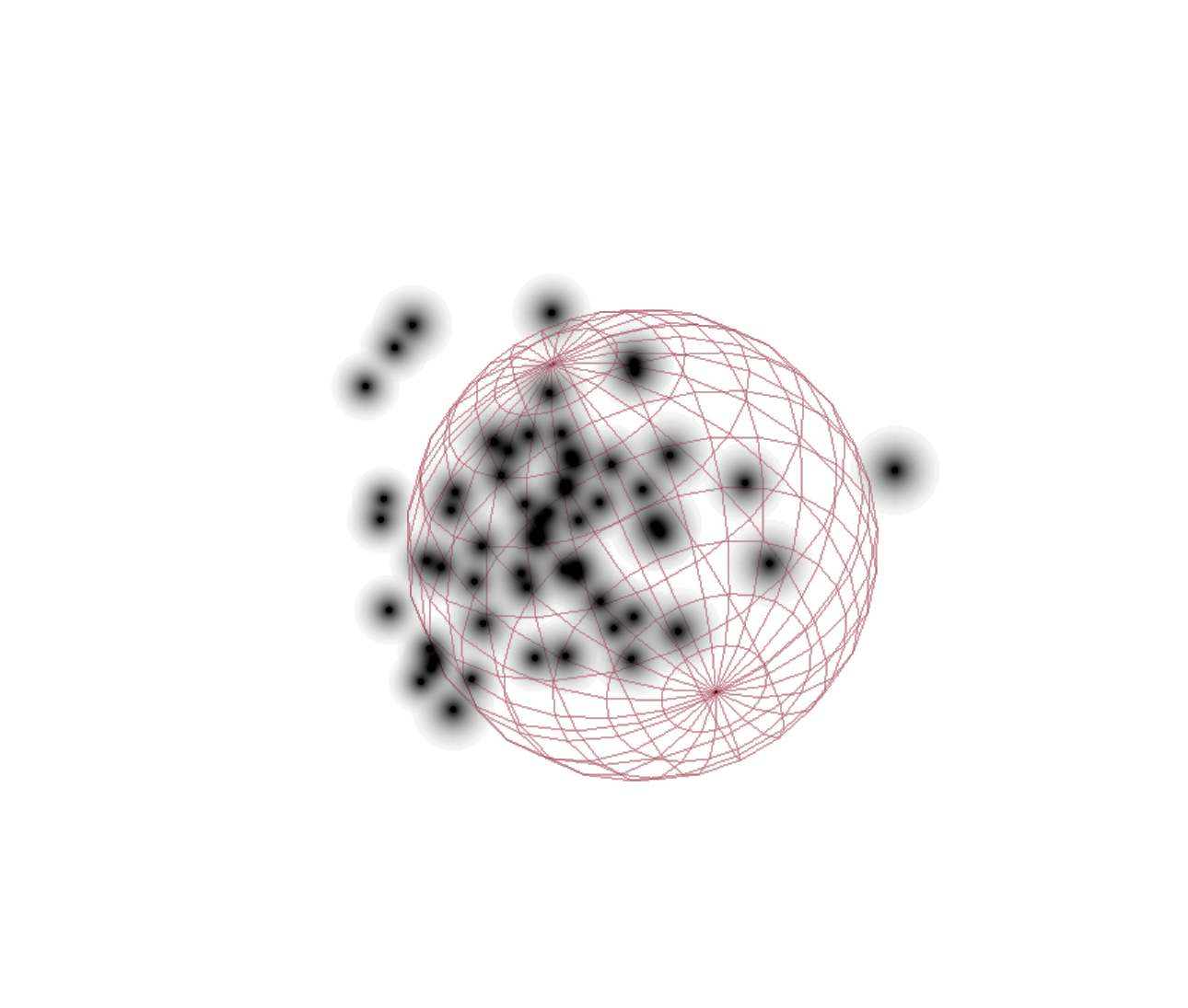}
    \includegraphics[width=0.49\textwidth]{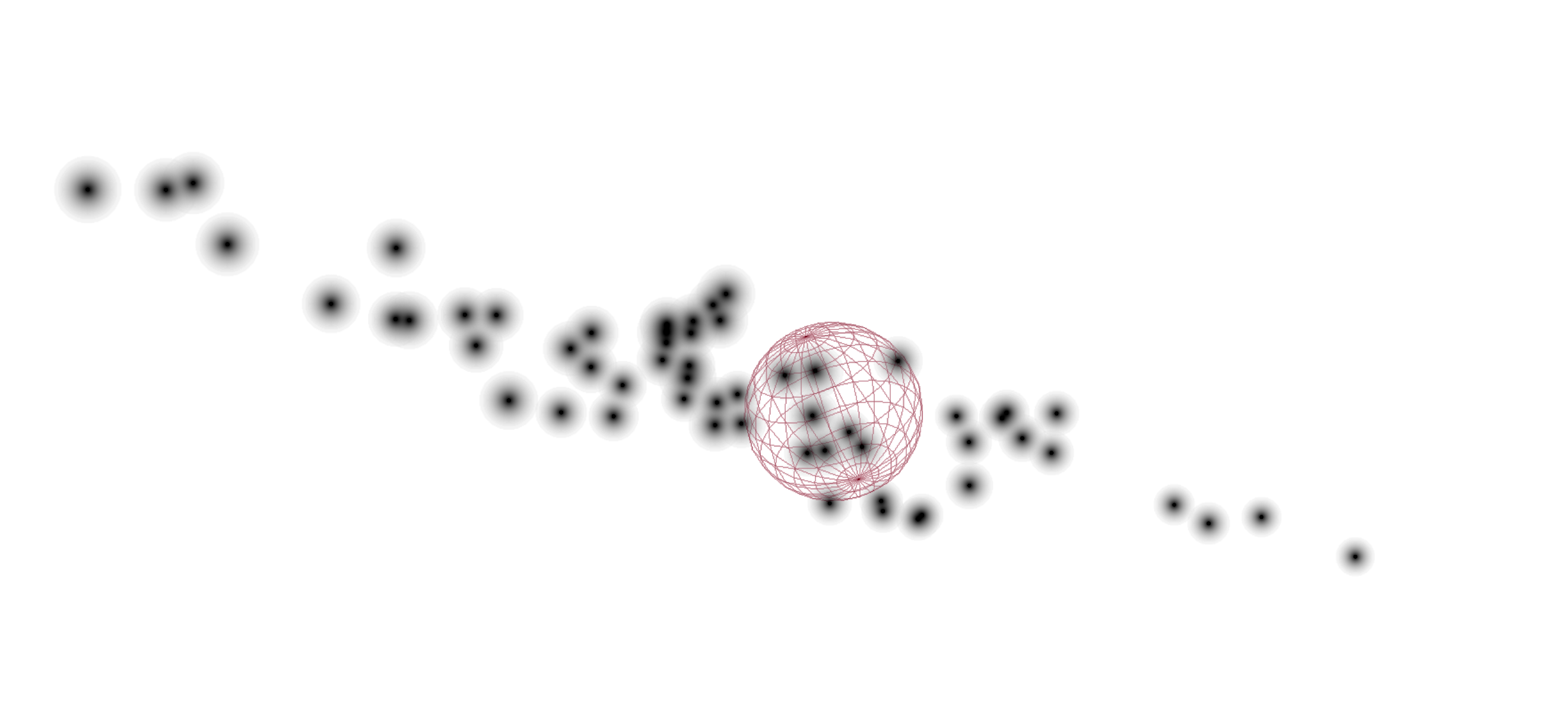}
    \caption{Example of a group identified in the TOA mock. The pink sphere shows the location of the group as found in the distorted mock (right), and is overlaid in the undistorted mock (left). This shows that the FoF algorithm finds physical groups whilst being run on a redshift distorted data set. A 3D video of this plot can be found \href{https://www.youtube.com/watch?v=8n5jkBoTX2g}{here}.}
    \label{fig:LinkValidation}
\end{figure*}

The mock was used to substantiate how accurately groups are recovered in the 2MRS given the relatively modest magnitude limit of $K_{\rm s}^{\rm o}<11\fm 75$ and the effect of incompleteness as a function of magnitude limit and observed recession velocity. The group finder was run on the TAO mock with a magnitude limit of $K_{\rm s}^{\rm o}<11\fm 75$. Groups with three members (representing the extreme case of low number statistics) were identified, providing the base for all future comparisons. The group finder was then run on increasingly deeper subsets of the mock in steps of $\Delta m = 0\fm25$ up to $K_{\rm s}^{\rm o}<14\fm 75$ (see x-axis of Fig.~\ref{fig:mock_plot}). 

The groups with three members found at $K_{\rm s}^{\rm o}<11 \fm 75$ were re-identified in subsequent, deeper runs of the group finder on the mock. Their positions (both on-sky and line-of-sight), the total number of members at each magnitude cut, and the difference of various group parameters compared to the $K_{\rm s}^{\rm o}<11 \fm 75$ limit were noted. These groups were then grouped into three distance bins of 50, 100 and 150~Mpc. Finally, the difference in the three-member group properties of the actual 2MRS were compared to the stepwise increased magnitude limit in the mock for each bin and are presented in the three panels of Fig~\ref{fig:mock_plot}.

The top panel in Fig.~\ref{fig:mock_plot} shows how the line-of-sight position of the group changes with deeper magnitude cuts. We note that line-of-sight positional offsets, i.e., the radial velocity axis, tend to increase with depth and slowly level off around 2 Mpc. This relatively large offset in line-of-sight position can be attributed to the well known ``finger-of-god'' effect. It never exceeds an Abell radius, though.

The middle panel shows the projected on-sky offset with depth. The average offset is less than 0.5 Mpc. This is much less than the error in the radial direction because there is no velocity distortion on the plane of the sky, other than the diminutive Kaiser flattening effect. Both the line-of-sight and on-sky offsets appear to be independent of the distance of the groups.

The bottom panel shows the percentage of the group present in a survey limited at $K_{\rm s}^{\rm o} \leq 11 \fm 75$. Unsurprisingly, there is a very clear trend with distance. Nearby groups are less affected by incompleteness than those at large distances. Thus, the fraction of recovered members is greater for nearby groups. We provide a correction to the group membership number based on the luminosity function of the relatively complete Virgo Supercluster, see \S4.2.4 below.

It is important to keep these offsets and incompleteness in mind when making use of this group catalogue.

\section{Visualization Techniques}
We made use of the state-of-the-art visualization laboratory hosted by the Institute of Data Intensive Astronomy (IDIA), which includes immersive displays and notably a Virtual Reality (VR) system that is ideal for 3D datasets. We were able to efficiently compare our group results (based on running the algorithm with our choice of linking length) against the intrinsic large-scale structure distribution provided by the mock. Figure~\ref{fig:LinkValidation} shows how well the algorithm recovers real groups using our choice of linking length parameter, by overlaying the groups that were found over the positions of the galaxies in the TAO mock.

\subsection{Visualization Facilities}
As mentioned, the catalogue was tested and verified using the new state-of-the-art visualization facilities hosted by IDIA. This included the use of Virtual Reality (VR), a panorama immersive facility, and the 8K digital planetarium dome housed at the Cape Town Iziko Museum. Using these facilities provides an additional supplement to the standard methods of bug-finding. This is achieved by overlaying several data sets, plus a variety of visual markers. This can consist of several other markers that can be used as diagnostic tools such as spheres (designating radius), lines connecting associated galaxies of the same group, colour palettes characterizing distance.

Every run of the algorithm and technical change made to the group finder was interrogated against previous renditions using the visualization lab, to inspect the results and ensure that: 1) no logical errors were present in the updated version, and 2) that changes rendered better results than previous iterations. 

\subsection{Rebuilding the Traditional FoF algorithm}
We first built a group-finder using the same method described in \cite{Huchra2007} and applied it to the early (shallower and incomplete) version of 2MRS. The two group catalogues were then compared using virtual reality in addition to running our new algorithm on this same data set. The galaxies from \cite{Huchra2007} were portrayed as small points in three dimensions using Cartesian Galactic coordinates. Both group catalogues were displayed as large blue and red spheres \citep[for][and our results, respectively]{Huchra2007}. The spheres mix colors in the virtual environment creating an overlap of purple (in this particular case). In this way it is possible to see at a glance (a) an agreement of the two catalogues (purple spheres) and (b) numerous types of discrepancies including line-of-sight offsets, on-sky offsets, groups found by our catalogue but not by \cite{Huchra2007}, and vice versa.

\subsection{Visualizing Graphs}
Using graph theory to average over the runs allowed us to think of each group as a topological data set, and as such they optimally lend themselves to visualization \citep{Pascucci2011Book}. This was accomplished by looking at pairs of galaxies, and deriving the frequency with which pairs were put in the same group. This results in a completely novel way of constructing a group catalogue, and solving the degeneracy issue arising from averaging over a three-dimensional dataset. 

Furthermore, we developed a unique approach in visualizing the multiple results by translating the mathematical graph objects from graph space into redshift space. Visual examination of the dataset then provides instantaneous feedback of both the underlying group finder and simultaneously the underlying mechanisms that led to those results. 

\begin{figure*}
    \centering
    \begin{tabular}{cc}
    \includegraphics[width=0.49\textwidth]{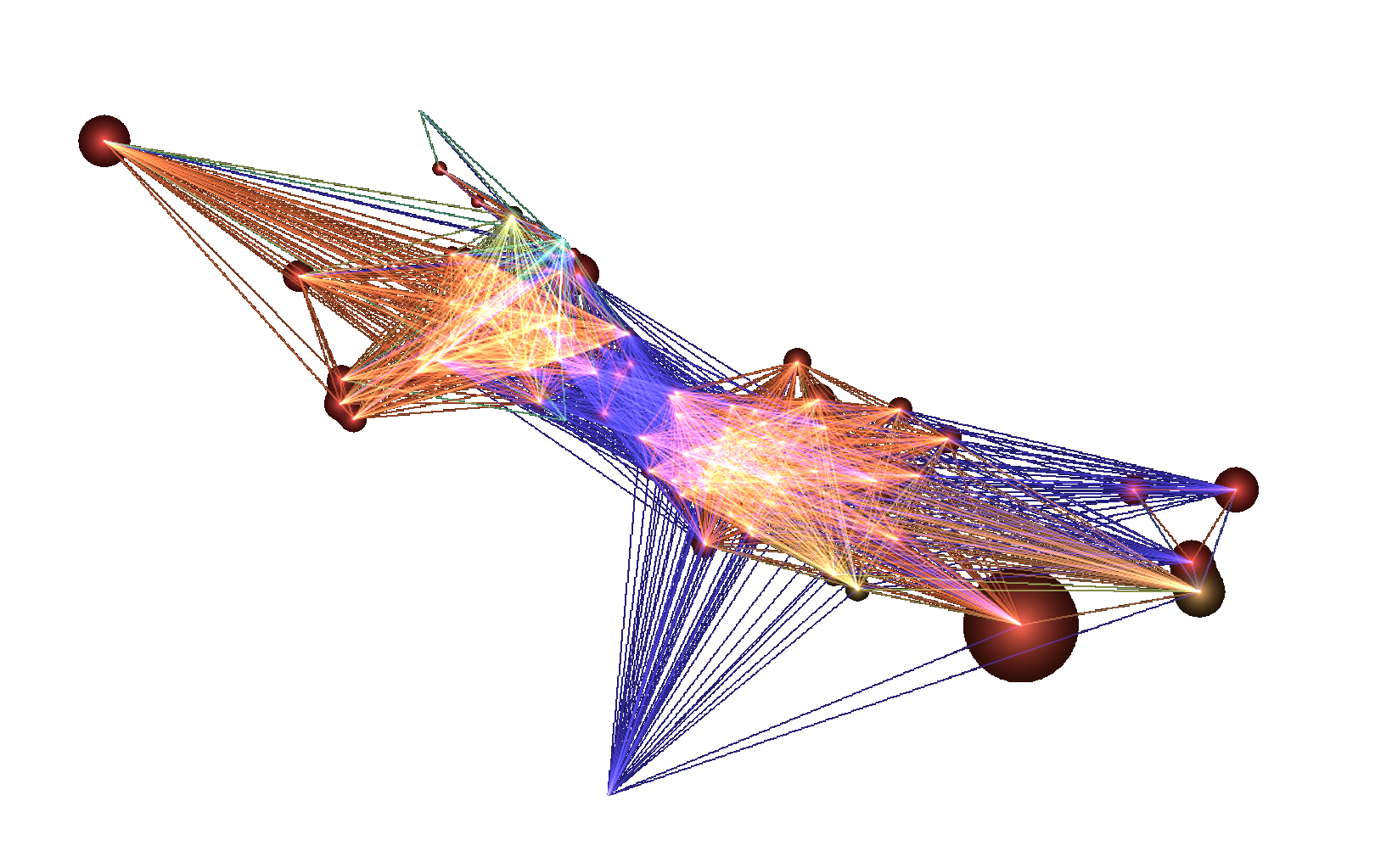} &
    \includegraphics[width=0.49\textwidth]{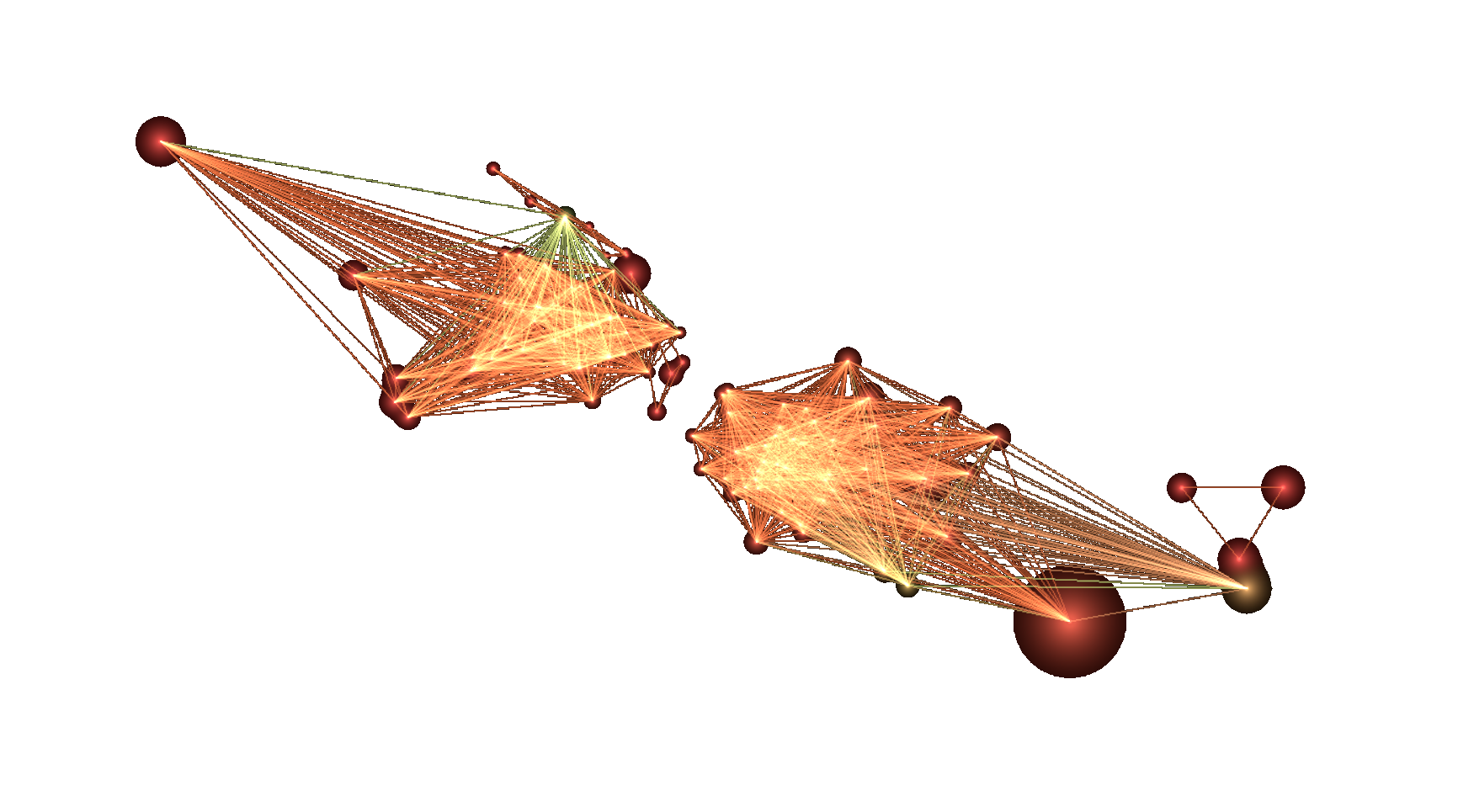} \\
    \textbf{(A)} & \textbf{(B)} \\
    \includegraphics[width=0.49\textwidth]{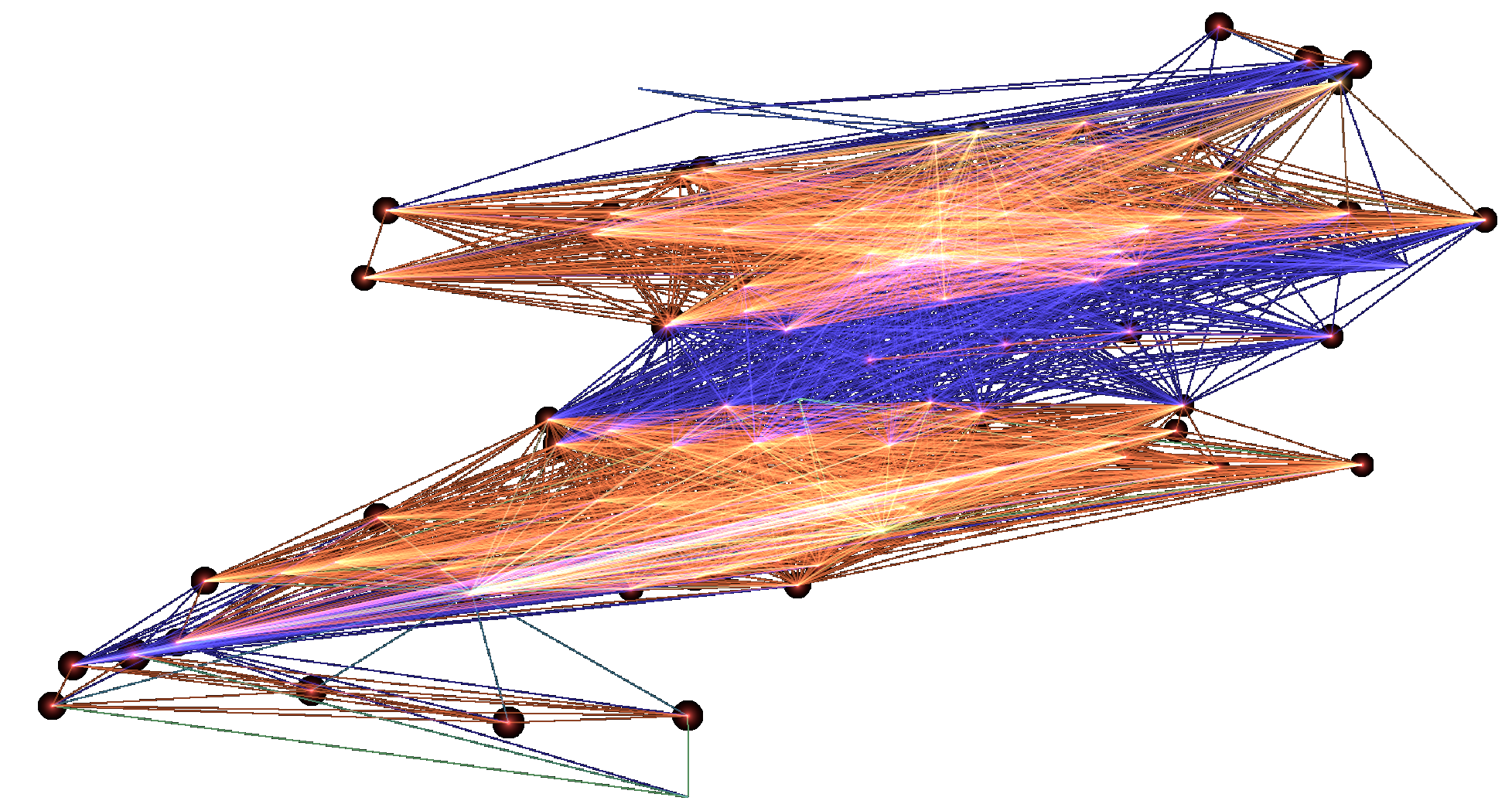} & 
    \includegraphics[width=0.49\textwidth]{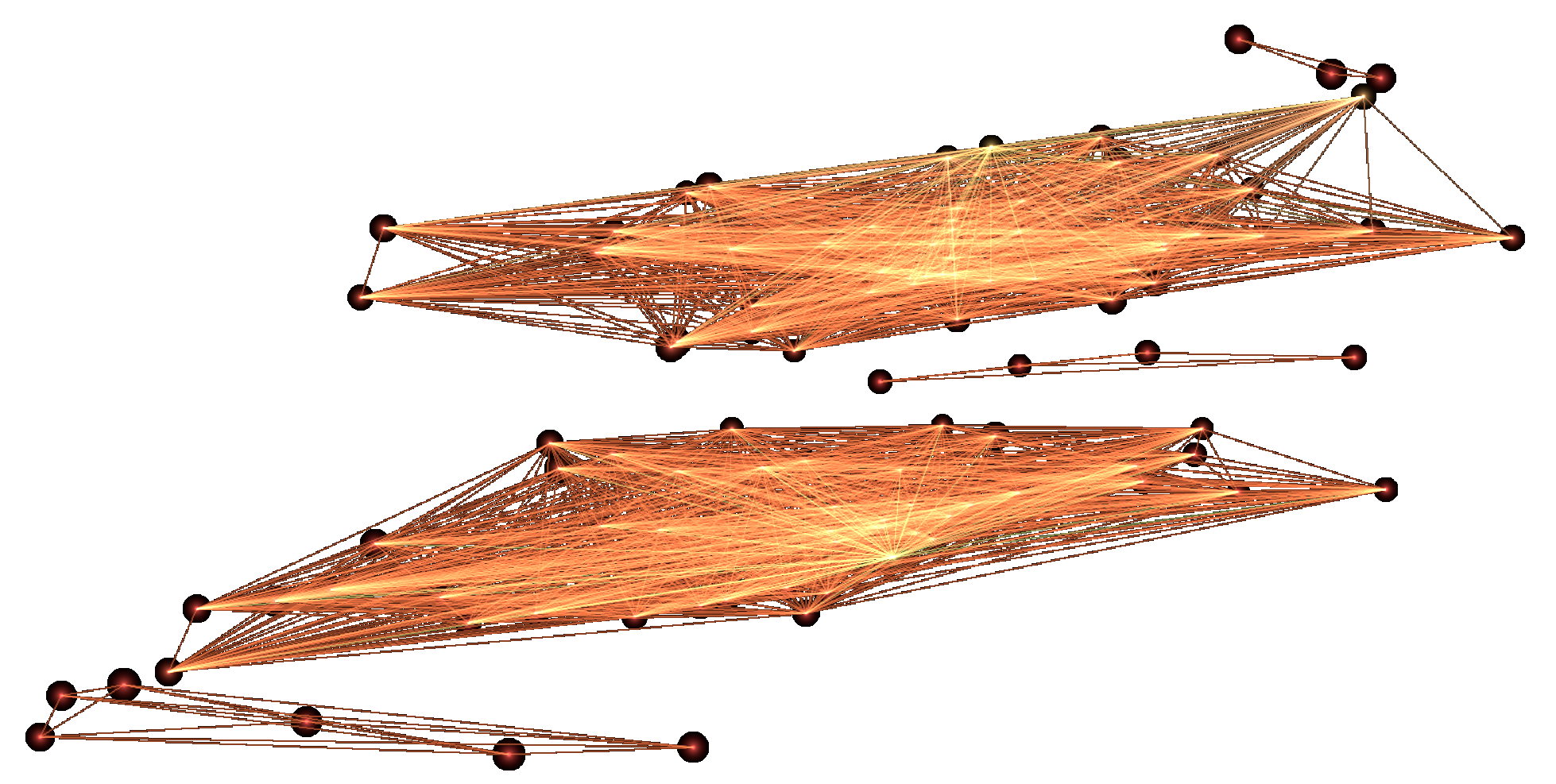} \\
    \textbf{(C)} & \textbf{(D)} \\

\end{tabular}
\caption{Three-dimensional example of a group with the resulting graph overlaid. Points represent the positions of 2MRS galaxies. Edges represent the number of connections of galaxies in the same group. Red edges represent 90$\%$ to 100$\%$ and blue edges represent $0 \%$ to $10\%$. A video displaying these connections in 3D is available \href{https://www.youtube.com/watch?v=tC2VFzs17WY}{here}.}
\label{fig:3DGraphexample}
\end{figure*}

Fig.~\ref{fig:3DGraphexample} shows an example of how the results of our group finder were visualized for any given group. Having the graphs displayed in this fashion allowed for exploration on how the graph structures themselves could be better used to improve the group results. This optimization was done by making use of: (1) the mock catalogue, which has both redshift and distance information; (2) well-known clusters and groups in the full 2MRS catalogue; and (3) any examples of groups which were found to be suspicious. The latter option was of particular importance, and possible only thanks to the powerful 3D visualization techniques. Finding examples of failures of the algorithm (e.g., where two obviously separate groups were connected as one) is straightforward when using visual inspection in VR and allows for exploration of intrinsic weaknesses of the group finder algorithm. 

\subsection{Removing Statistical Outliers}
Visual inspection reveals that averaging over all results is not good enough. While the robustness of the method is assured, groups can become too large because of a single connection/edge between two obviously distinct groups (eg., see Fig \ref{fig:Bad_Virgo}). It also is unrealistic to accept every single connection. A galaxy which is associated with a group for $10\%$ of the runs but with another one for the remaining $90\%$ is much more likely to form part of the latter. In other words, while the robustness of the method is assured, robustness to statistical outliers is not and a significance cut must be made. A careful analysis based on both the mock catalogue and well-known groups and clusters indicated that accepting connections at the $>50\%$ level occurrence gives the best results. Using the graph tool was highly effective here; a removal of all edges with a weighting $<50\%$ was sufficient to disconnect subgraphs. At this point, the group finder is {stable}. An example of this process is shown in Fig.~\ref{fig:3DGraphexample}. Edges are coloured according to the percentage of runs in which a galaxy pair was assigned to the same group. Frames (A) and (B) display the galaxy group on the plane of the sky while (C) and (D) show the distribution along the line-of-sight. Frames (A) and (C) show the group before the cut was implemented, while frames (B) and (D) show the effect of the cut. Note that if the cut had not been implemented (i.e. removal of low-weighted, blue edges) then the entire system would have been classified as a single group.  However, confusion can sometimes occur in situations where groups are radially aligned coincidentally.

\subsection{Subgroups}
On the other hand, we also came across several groups that seemed to have two or more concentrations of galaxies. If the statistical cut had been stricter than $50\%$ (say $60\%$ or $70\%$) then they would have split into separate groups. To account for these cases, a higher statistical cut was applied to each group to check the likelihood of them having real substructure, and/or whether some of these subgroups are real. Differentiation of these cases was easy in VR; however, translating this visual information into an algorithmic equivalent is quite a challenge. The connectedness score described in Eq.~\ref{eq:Score} achieved this goal --  coloring the galaxies by this property made the groups that required attention stand out clearly. Based on this information, we could identify groups with low average connectedness score and which also contained subgroups. In these cases, these subgroups became their own groups. 

\begin{figure}

    \centering
\begin{tabular}{c}
    \includegraphics[width=0.49\textwidth]{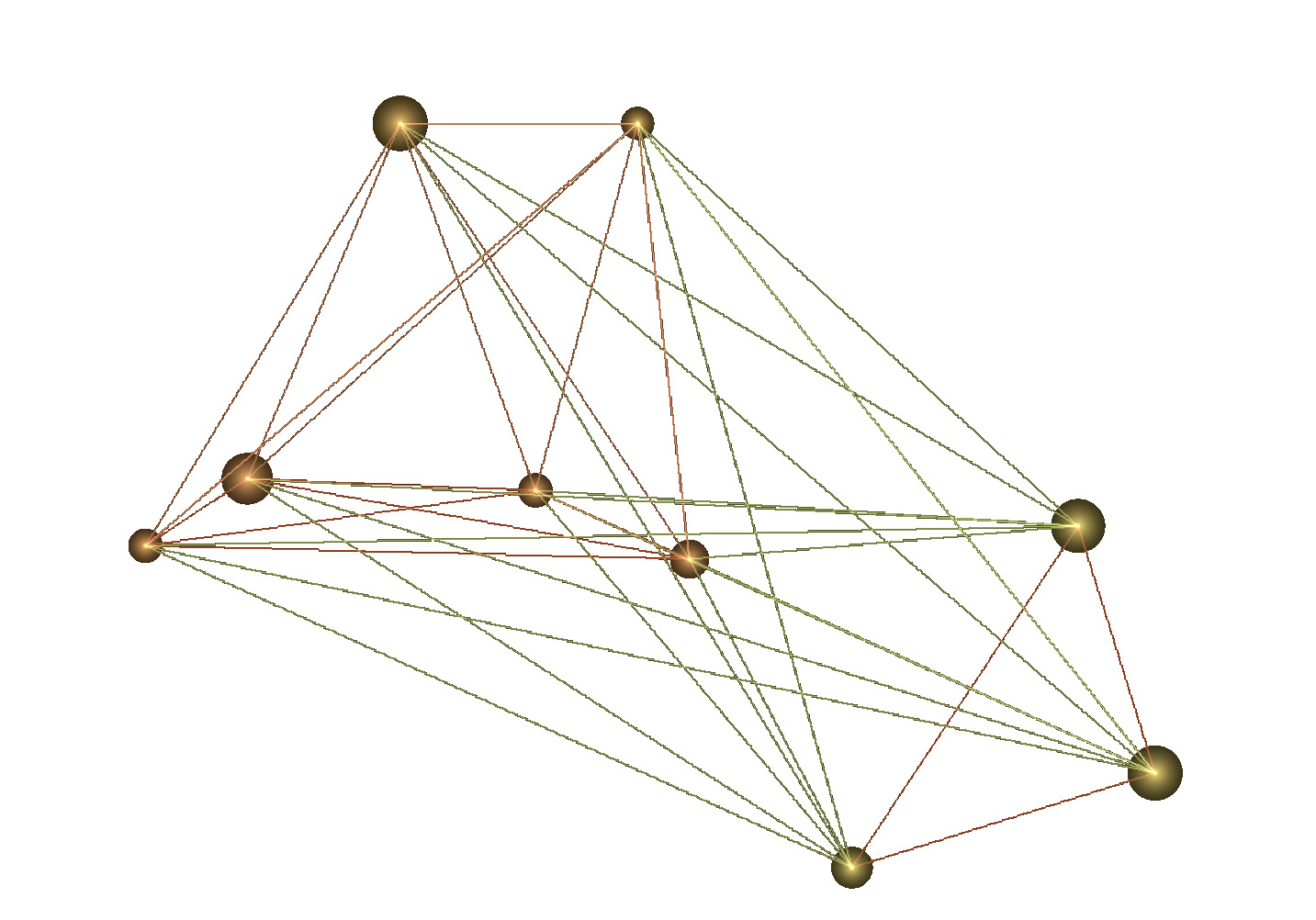} \\
    \textbf{(A)} \\

    \includegraphics[width=0.49\textwidth]{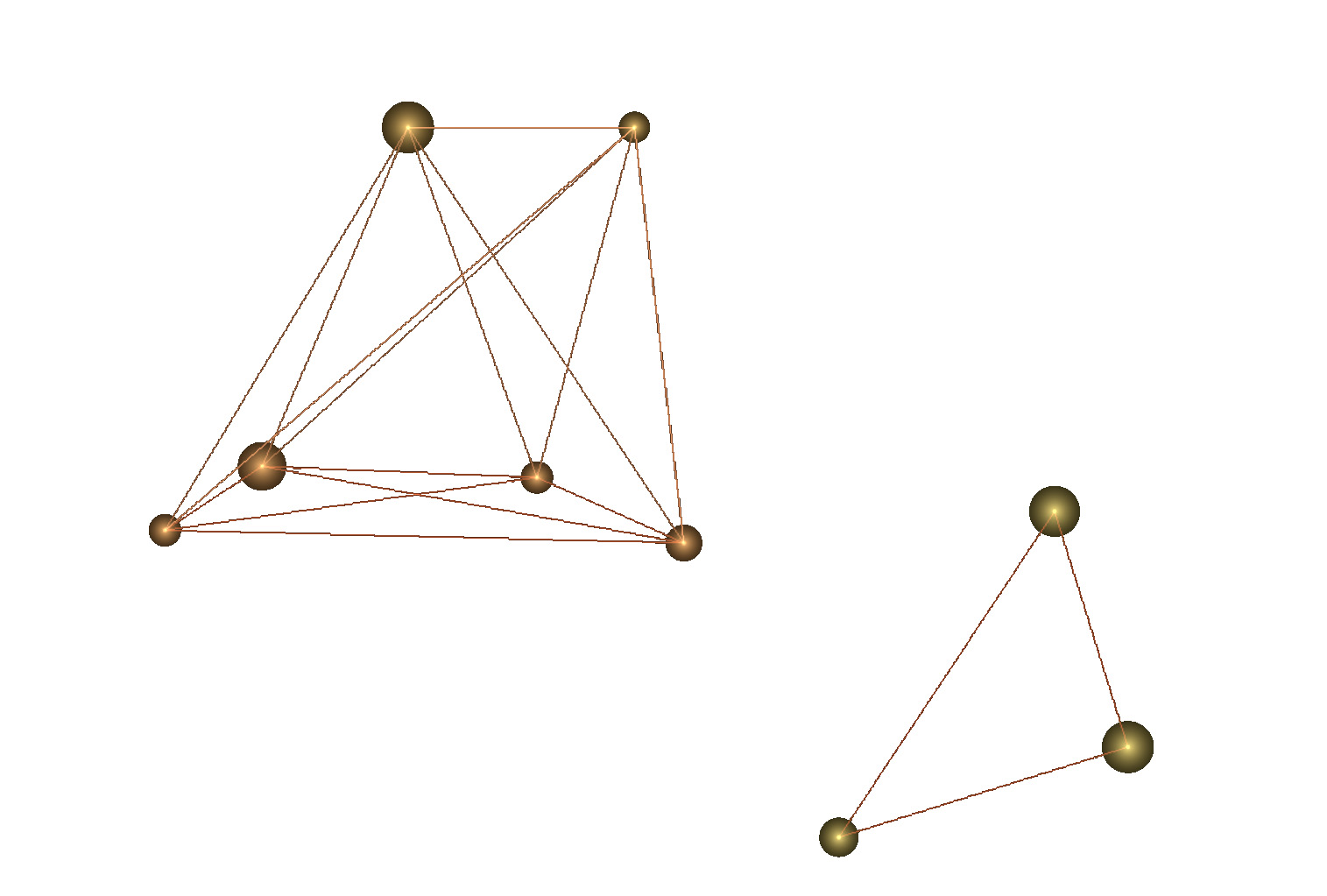} \\
    \textbf{(B)} \\
    
    \includegraphics[width=0.49\textwidth]{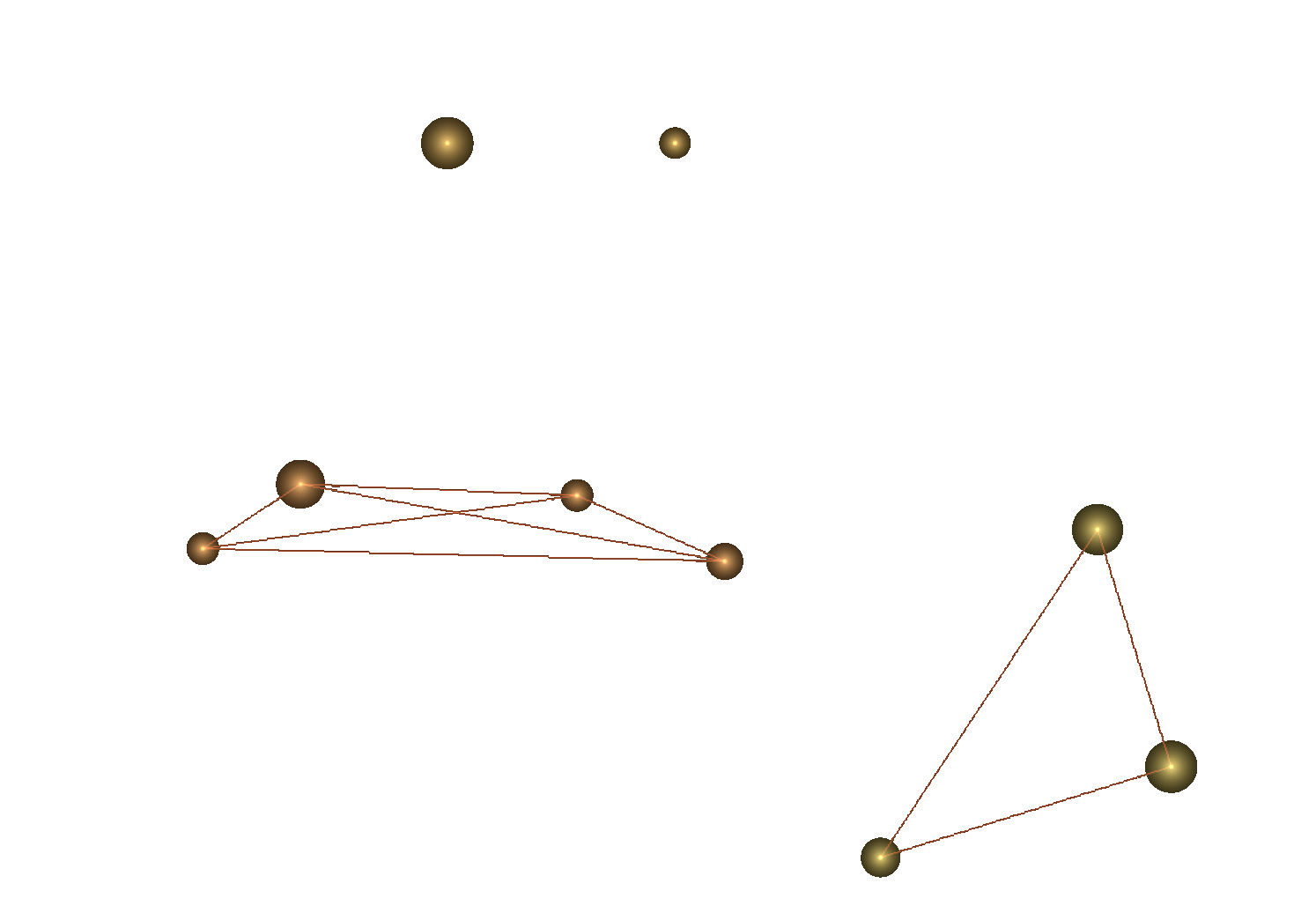} \\
    \textbf{(C)}
\end{tabular}

\caption{Example of a galaxy group containing two subgroups and highlighting the method of identifying subgroups by incrementally removing the weakest edges. Panel (A) shows the group with all edges $\geq 50\%$  still present, panel (B) shows the group after edges with a weighting of $60\%$ and less are removed, and  panel (C) shows the group when edges with a weighting of $80\%$ and less are removed, resulting in only the strongest edges remaining (in this case edges with a weighting of 1).  A 3D-animation of these examples is available \href{https://www.youtube.com/watch?v=6Al0SRB-fCs}{here}.}
\label{fig:SubGroups}

\end{figure}

Fig.~\ref{fig:SubGroups} gives an illustration of a group with subgroups. From frame (A) to (C), lines of decreasing strengths are removed until no more lines can be cut. The number of separate structures changes from one in frame (A) to two in frames (B) and (C). Because the maximum number of isolated systems is two, subgroups are chosen to be the first instance at which the maximum number of isolated systems occurs (frame (B) in this particular example). 

The use of numerous visualization techniques enabled us to identify issues affecting the traditional FoF algorithm and make various updates to address them. These included: (1) running the algorithm many times; (2) averaging over all runs using graph-theory; (3) removing statistical outliers by implementing a significance cut; (4) applying a recursive cut to graphs to find any subgroups; and (5) identifying weak and nonphysical connections and correcting these groups.  

\section{The 2MRS Group catalogue}
Our final 2MRS group catalogue contains 3022 entries and is the most complete ``whole-sky'' galaxy group catalogue to date. We provide measured and calculated group properties, such as: 3D positions, luminosity and comoving distances, observed and corrected number of members, richness metric, velocity dispersion, and estimates of $R_{200}$ and $M_{200}$.

\subsection{Measured Group Properties}
The measured group properties include the velocity dispersion and the projected central position, which in turn affect many of the other derived quantities. Thus, we elaborate on these below.

\subsubsection{Velocity dispersion}
We use two methods for calculating the velocity dispersion, namely the standard deviation and the root mean square (RMS) of the peculiar velocity distribution along the line-of-sight, described as 
\begin{equation}
    v_i=c\frac{z_i-\bar{z}}{1+\bar{z}}
\label{eq:TomDisp}
\end{equation}
\noindent {where ${z}_i$,} $\bar{z}$ are the velocity of galaxy $i$ and the mean redshift of the galaxy group, respectively \citep{Navarro1995,Jarrett2017}.

\subsubsection{Projected group center}
There have been several proposed methods of determining the projected group center amongst different surveys and catalogues \citep{Huchra2007,Poggianti2010,Robotham2011,Tempel2018}. Several techniques were explored using the TAO mock catalogue, such as: luminosity-weighted and flux-weighted moments, and the simple geometric center. We found that taking the average RA and Dec of the group members rendered the most {accurate} central on-sky position, both when inspecting the mock catalogue and identifying well-known existing groups and clusters. 

\subsubsection{Projected Radius}
 The projected radius was calculated as the maximum projected on-sky distance of all the galaxies from the projected group center. While this method is not robust against outliers, we found that those could be robustly identified and corrected using the new visualization techniques.
 
 The projected radius was calculated as the maximum projected on-sky distance of all the galaxies from the projected group center. While this method is not robust against outliers, it was found that those could be robustly identified and corrected using the new visualization techniques {as visual spheres with radii equal to the projected radius metric would become hugely exaggerated, immediately highlighting problematic groups and methods which were then ultimately corrected, resulting in the complete catalogue having no such issues.}

\subsection{Calculated Group Properties}
\subsubsection{$R_{200}$ and $M_{200}$}
Key properties in many group catalogues are $R_{200}$ and $M_{200}$, defined as the radius of the sphere with 200 times the critical density and the total mass enclosed within said sphere, respectively. Assuming the virial theorem, $R_{200}$ and $M_{200}$ can be calculated as follows \citep{Poggianti2010}: 
\begin{equation}
    R_{200}=2.37 \sigma \left( {\Omega_{\Lambda} + \Omega_M \left( 1+z\right)^3} \right )^{-1/2} \text{ Mpc}\label{eq:R200}
\end{equation}
\begin{equation}
    M_{200}=1.6 \times 10^{15} \sigma^3 \left ( {\Omega_{\Lambda}+\Omega_M\left(1+z\right)^3}\right )^{-1/2} \text{ M}_{\odot}\label{eq:M200}
\end{equation}
where $\sigma$ is the line-of-sight velocity dispersion, expressed in units of $10^3\text{ km\,s}^{-1}$ and ${H_{0} = 73}$ km s$^{-1}$ Mpc$^{-1}$.

\subsubsection{Comoving and Luminosity Distances}
Comoving and Luminosity distances, as described by \cite{Peebles1993Book}, are calculated as
\begin{equation}
    d_c=\frac{c}{H_0}\int\limits_0^z {d z^{\prime}}\left ( {\Omega_M (1+z^{\prime})^3 + \Omega_k (1+z^{\prime})^2 + \Omega_{\Lambda}} \right )^{-1/2}
\label{eq:CoMoving}
\end{equation}
\noindent and $d_L=d_c(1+z)$, respectively. 
\subsubsection{Richness}
A simple definition of richness similar to \cite{Andreon2016} is included in the catalogue. We define richness as the number of galaxies within a group with absolute magnitude of $M_K^{\rm o}<-23\fm 5$. This definition is maintained up to 100 Mpc where the survey is still reasonably complete. The richness is still calculated for groups greater than 100 Mpc but this value would represent the minimum richness possible and is reported as such in the final catalogue. 

\subsubsection{Corrected Group Members}
We provide a simple correction to the number of members in a group in order to gain a relative estimate of the true number of members were there no incompleteness. This is done by first assuming that the Virgo Supercluster is complete. The number of corrected group members for a given group ($N_{\text{mem}}'$) can then be calculated as
\begin{equation}
   N_{\text{mem}}' =  N_{\text{mem}} \left(\frac{N_{\rm v}}{N_{\rm v}(d)}\right)
\label{eq:correction}
\end{equation}
where $N_{\text{mem}}$ and $N_{\rm v}$ are the number of members found in the given group and the number of members found in Virgo, respectively. $N_{\rm v}(d)$ is the number of group members that Virgo would have if it were at the same distance ($d$) of the given group. This is strictly dependent on the luminosity function of Virgo. 

\subsection{Catalogues}
We present one main and two supplementary 2MRS group catalogues, as detailed below.

\subsubsection{The 2MRS catalogue}
The main 2MRS group catalogue is presented in Table \ref{tbl:catalogue}. The 20 groups with the highest membership are shown for explanatory purposes, while the full table is available online. The columns are:
\\\\
\noindent(1) - 2MRS group ID as defined in this work. \\ 
(2) - Names (where available) of well-known groups and clusters in other catalogues. \\
(3) - Average RA ($\alpha$) of the group, in degrees.\\ 
(4) - Average Dec ($\delta$) of group, in degrees.\\ 
(5) - Average Galactic longitude ($l$) of the group, in degrees.\\ 
(6) - Average Galactic latitude ($b$) of the group, in degrees.\\ 
(7) - Number of 2MRS galaxies in the group. \\ 
(8) - Corrected number of members, calculated using Eq.~\ref{eq:correction}.\\
(9) - Richness, as defined in \S4.2.3\\
(10) - Recession velocity, in the CMB reference frame. \\ 
(11) - Co-moving distance, calculated using Eq.~\ref{eq:CoMoving}. \\ 
(12) - Luminosity distance, as defined in \S4.2.2. \\
(13) - Velocity dispersion, calculated using Eq.~\ref{eq:TomDisp} \\
(14) - Velocity dispersion given by the standard deviation of the velocities of members of the group. \\
(15) - $R_{200}$ value, calculated using Eq.~\ref{eq:R200}. \\ 
(16) - $M_{200}$ value, calculated using Eq.~\ref{eq:M200}. \\
(17) - Number of subgroups, as defined in \S2.3.\\ 

\begin{figure}
    \centering
    \includegraphics[width=0.5\textwidth]{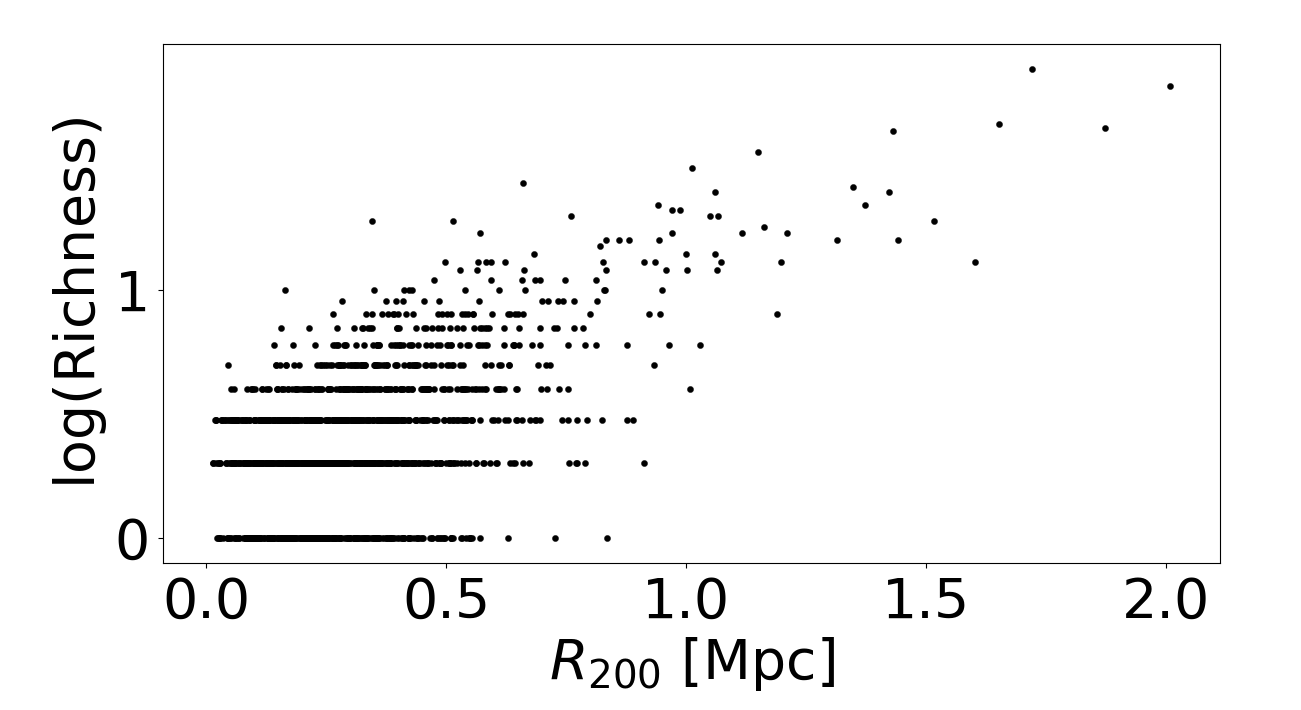}
        \caption{Relation between group richness (\S4.2.3) and $R_{200}$.}
    \label{fig:R200vRichness}
\end{figure}

\begin{figure}
    \centering
    \includegraphics[width=0.5\textwidth]{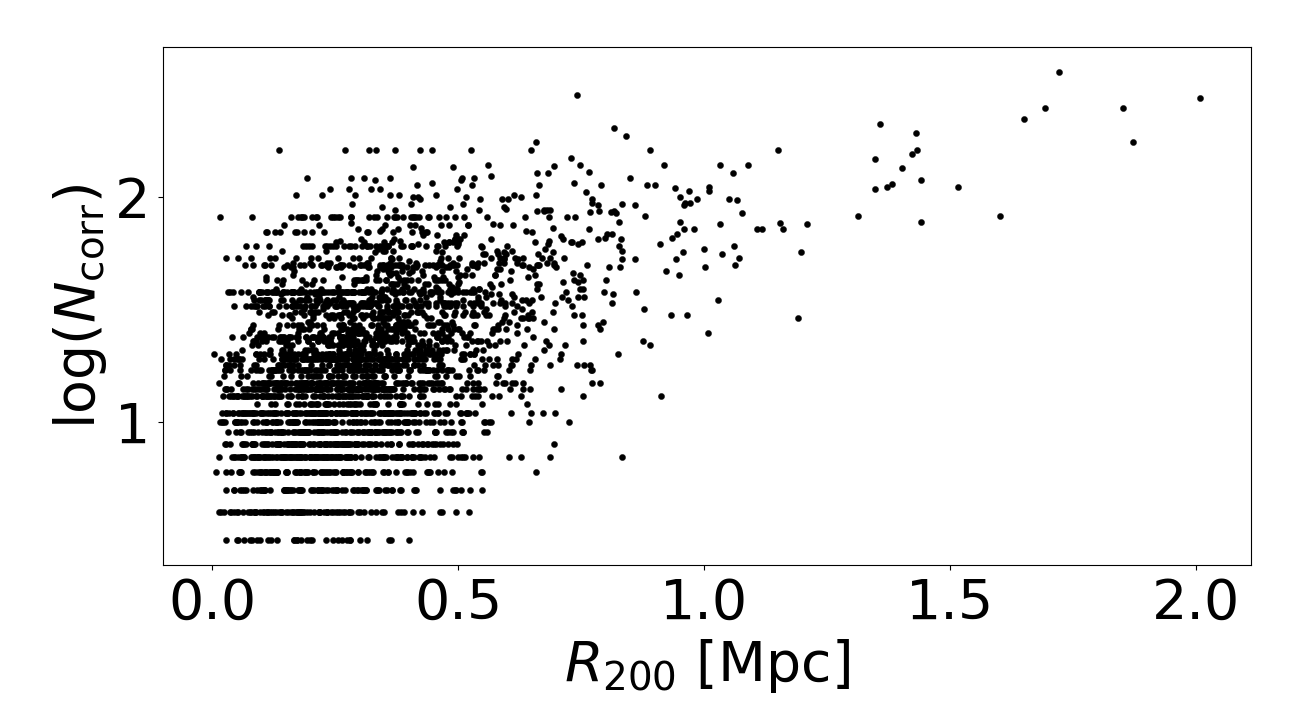}    
    \caption{Relation between number of corrected group members and $R_{200}$.}
    \label{fig:R200vNmem}
\end{figure}

We have plotted both richness (column 9 in Table~\ref{tbl:catalogue}) and the number of corrected members, $N^{'}_{\rm mem}$ (column 7 in Table~\ref{tbl:catalogue}) against $R_{200}$ in Fig.~\ref{fig:R200vRichness} and Fig.~\ref{fig:R200vNmem}, respectively. $R_{200}$ is determined solely by the dispersion of the group, unlike richness and $N_{\rm mem}$ which are entirely independent of dispersion. It therefore is necessary to have both the number of members and richness increase with $R_{200}$. Our definition of richness also shows a tight relation with $R_{200}$ which is a good validation of our procedure. The extreme outlier present in Fig.~\ref{fig:R200vNmem} is the Virgo Supercluster. Figures~\ref{fig:R200vRichness} and \ref{fig:R200vNmem} show the integrity of the main catalogue.  

\subsubsection{The 2MRS Subgroup catalogue}
A subgroup catalogue is also included, with a random sub-sample shown in Table~\ref{tbl:SubGroups}. The subgroup catalogue includes Group ID, mean values of $\alpha$, $\delta$, $l$ and $b$, $v_{\text{cmb}}$, and $N_{\text{mem}}$ as described in Table \ref{tbl:catalogue}. The sub ID, (1) in Table \ref{tbl:SubGroups}, demarcates the unique identifier for each individual subgroup. The ID is made up of two parts, namely the host galaxy group ID (the group to which said subgroup belongs) and the numbered subgroup within said host group. 

\subsubsection{The 2MRS Galaxies in Groups catalogue}
We also include a catalogue consisting of all the galaxies that are member of a group in Table \ref{tbl:groupmembers}. This catalogue includes the 2MASS ID, $\alpha$, $\delta$, $l$, $b$ of each galaxy as given in \cite{Macri2019}, as well as our calculated $v_{\text{cmb}}$, Group ID from Table \ref{tbl:catalogue} and Sub ID from Table \ref{tbl:SubGroups}. A random sub-sample of galaxies are shown in Table \ref{tbl:groupmembers} for reference.  

The on-sky distribution of the final 2MRS group catalogue is shown in Fig.~\ref{fig:FamousAitoffGroup}. We have recovered all the well-known, large-scale structures and, in particular, the intricate substructures which comprise them. This bodes well in validating our group finding algorithm as the structures we recovered are similar to previous 2MRS-based catalogues \citep{Huchra2007,Saulder2016,Tempel2016,Kourkchi,Tempel2018}. 

\subsection{Comparisons to Literature}
In validating our results, it is important to compare our work with previous catalogues based on earlier versions of 2MRS, and to compare our method with other techniques. For the former, we compared our catalogue against \cite{Huchra2007}, \cite{Tempel2016}, \cite{Lim2017} and \cite{Tempel2018}. Our cross-matching search was done using tolerances of 1 and 5 Mpc for plane-of-sky and line-of-sight distances, respectively. We choose these values to be conservative and expect matches to be well within 1 Mpc in the spatial plane and 5 Mpc in the radial. The resulting matches and discrepancies were investigated using the IDIA visualization lab.  

\subsubsection{Crook et al., (2007)}
\cite{Huchra2007} based their group finder on the classic FoF algorithm as developed by \cite{Huchra1982} without any modification. This group finder was applied to an earlier and shallower 2MRS catalogue with a limit of $K_{\rm s}^{\rm o} < 11\fm25$. Nevertheless, since we built upon their techniques for our analysis, a comparison is appropriate.

\cite{Huchra2007} used two different choices of linking length, resulting in two catalogues (respectively referred to as high- and low-density). We only compared our results against the high-density catalogue since it used the same linking lengths proposed by \cite{Ramella1997} and adopted by us. We did not consider any groups found in this previous work that were based on artificial galaxies, which they used to fill the ZoA.

\vfill\pagebreak\newpage

\begin{sidewaystable*}
\vspace{-16cm}
\caption{2MRS Galaxy Group catalogue}
\resizebox{\linewidth}{!}{%
\begin{tabular}{rcS[table-format=3.2]S[table-format=3.2]S[table-format=3.2]S[table-format=3.2]ccrcS[table-format=3.2]S[table-format=3.2]ccccc}
\hline \hline \\
Group ID & Other Names      & \multicolumn{1}{c}{$\alpha$}  & \multicolumn{1}{c}{$\delta$} & \multicolumn{1}{c}{$l$}       & \multicolumn{1}{c}{$b$}       & $N_{\text{mem}}$ & $N'_{\text{mem}}$ & \multicolumn{1}{c}{$R$} & $v_{\text{cmb}}$  & \multicolumn{1}{c}{$d_c$}     & \multicolumn{1}{c}{$d_L$}     & RMS               & $\sigma$          & $R_{200}$ & $M_{200}$                        & $N_{\text{subs}}$  
\\ \\
         &                  & \multicolumn{4}{c}{{[}deg, J2000{]}} &  &     &            & {[}km~s$^{-1}${]} & \multicolumn{2}{c}{{[}Mpc{]}} & \multicolumn{2}{c}{{[}km~s$^{-1}${]}} & {[}Mpc{]} & {[}$\times 10^{11} M_{\odot}${]} &       \\ \\ 
         
(1)      & (2)              & \multicolumn{1}{c}{(3)}       & \multicolumn{1}{c}{(4)}       & \multicolumn{1}{c}{(5)}       & \multicolumn{1}{c}{(6)}       & (7) & (8)   & \multicolumn{1}{c}{(9)}      & (10)              & \multicolumn{1}{c}{(11)}      & \multicolumn{1}{c}{(12)}      & (13)              & (14)              & (15)      & (16)                             & (17)  \\ \\         
         \hline \hline
2987     & VIRGO CLUSTER    & 187.80                         & 11.41                         & 285.13                        & 73.58                         & 163 & 163   & 36                           & 1523              & 20.8                          & 20.9                          & 486               & 484               & 1.15      & 188.5                            & 0     \\
153      & ABELL 3627       & 243.67                        & -60.89                        & 325.27                        & -7.17                         & 121 & 278   & 67                           & 4931              & 67.3                          & 68.4                          & 854               & 861               & 2.01      & 1015.6                           & 3     \\
50       & ABELL 3526B      & 192.03                        & -41.21                        & 302.26                        & 21.66                         & 102 & 177   & 45                           & 3784              & 51.7                          & 52.3                          & 795               & 792               & 1.87      & 820.8                            & 3     \\
3031     & ABELL 0426       & 49.76                         & 41.38                         & 150.53                        & -13.45                        & 95  & 224   & 47                           & 5158              & 70.4                          & 71.6                          & 702               & 714               & 1.65      & 564.4                            & 0     \\
228      & COMA CLUSTER     & 194.90                         & 28.00                          & 59.10                          & 87.99                         & 95  & 360   & 78                           & 7260              & 98.9                          & 101.3                         & 734               & 751               & 1.72      & 643.8                            & 3     \\
179      & -                & 159.12                        & -27.66                        & 269.64                        & 26.35                         & 85  & 156   & 25                           & 4043              & 55.2                          & 56.0                          & 604               & 611               & 1.42      & 360.2                            & 0     \\
271      & ABELL 0262       & 28.44                         & 36.23                         & 136.75                        & -24.96                        & 61  & 129   & 25                           & 4630              & 63.2                          & 64.2                          & 450               & 457               & 1.06      & 148.7                            & 0     \\
3011     & -                & 181.50                         & 46.19                         & 145.21                        & 68.92                         & 61  & 61    & 4                            & 993               & 13.6                          & 13.6                          & 193               & 193               & 0.46      & 11.9                             & 0     \\
500      & ABELL 1367       & 176.15                        & 19.97                         & 234.45                        & 73.12                         & 56  & 194   & 44                           & 6787              & 92.5                          & 94.6                          & 610               & 623               & 1.43      & 368.4                            & 0     \\
151      & ABELL S0805      & 281.49                        & -63.03                        & 332.53                        & -23.38                        & 56  & 111   & 31                           & 4392              & 60.0                          & 60.8                          & 429               & 424               & 1.01      & 129.2                            & 0     \\
25       & ERIDANUS CLUSTER & 54.21                         & -20.12                        & 211.53                        & -51.64                        & 53  & 53    & 5                            & 1517              & 20.8                          & 20.9                          & 251               & 250               & 0.59      & 26.1                             & 0     \\
84       & HYDRA CLUSTER    & 157.79                        & -35.36                        & 273.17                        & 19.27                         & 51  & 76    & 17                           & 3155              & 43.1                          & 43.6                          & 513               & 517               & 1.21      & 221.0                            & 2     \\
4        & FORNAX CLUSTER   & 53.93                         & -35.02                        & 236.00                         & -54.21                        & 49  & 49    & 9                            & 1357              & 18.6                          & 18.7                          & 296               & 295               & 0.70       & 42.3                             & 0     \\
202      & -                & 17.31                         & 32.65                         & 127.26                        & -30.08                        & 48  & 107   & 21                           & 4838              & 66.0                          & 67.1                          & 412               & 417               & 0.97      & 114.3                            & 2     \\
150      & -                & 193.67                        & -13.77                        & 304.14                        & 49.10                          & 48  & 106   & 20                           & 4718              & 64.4                          & 65.4                          & 323               & 328               & 0.76      & 54.8                             & 3     \\
369      & OPH CLUSTER      & 258.06                        & -23.39                        & 0.53                          & 9.30                           & 46  & 250   & $\geq$ 46                      & 8757              & 119.2                         & 122.6                         & 792               & 812               & 1.85      & 804.5                            & 2     \\
101      & -                & 20.83                         & 33.54                         & 130.51                        & -28.86                        & 46  & 97    & 20                           & 4614              & 63.0                          & 64.0                          & 453               & 460               & 1.07      & 151.8                            & 0     \\
261      & -                & 193.35                        & -9.26                         & 303.75                        & 53.61                         & 45  & 92    & 16                           & 4473              & 61.1                          & 62.0                          & 365               & 369               & 0.86      & 79.4                             & 2     \\
355      & -                & 107.01                        & 49.74                         & 167.35                        & 22.93                         & 44  & 128   & 27                           & 6032              & 82.2                          & 83.9                          & 281               & 286               & 0.66      & 36.1                             & 3     \\
142      & -                & 206.98                        & -30.39                        & 317.13                        & 30.92                         & 44  & 98    & 21                           & 4853              & 66.2                          & 67.3                          & 419               & 424               & 0.99      & 120.2                            & 0     \\ \hline
\end{tabular}}
\label{tbl:catalogue}
\end{sidewaystable*}

\begin{table*}
\caption{2MRS SubGroup catalogue}
\begin{tabular}{rrd{2}d{2}d{2}d{2}rr}
\hline \hline \\
Sub ID & Group ID & \multicolumn{1}{c}{$\alpha$}  & \multicolumn{1}{c}{$\delta$} & \multicolumn{1}{c}{$l$}    & \multicolumn{1}{c}{$b$}       & \multicolumn{1}{c}{$v_{\rm cmb}$} & $N_{\rm mem}$ \\ \\
       &                & \multicolumn{4}{c}{{[}deg, J2000{]}} & \multicolumn{1}{c}{[km~s$^{-1}$]}      &      \\ \\ \hline \hline
1000-1 & 1000          & 140.72                        & 24.44                         & 203.96                     & 43.31                         & 10337                                & 4    \\
1000-2 & 1000          & 141.60                         & 23.89                         & 204.98                     & 43.94                         & 10145                                & 3    \\
1015-1 & 1015          & 107.14                        & -49.10                         & 259.72                     & -17.53                        & 12790                                & 7    \\
1015-2 & 1015          & 106.12                        & -48.98                        & 259.35                     & -18.11                        & 12997                                & 3    \\
1031-1 & 1031          & 241.04                        & 69.77                         & 103.49                     & 39.28                         & 7540                                 & 3    \\
1031-2 & 1031          & 239.39                        & 70.74                         & 104.93                     & 39.22                         & 7486                                 & 6    \\
1091-1 & 1091          & 52.65                         & 41.59                         & 152.24                     & -12.05                        & 5367                                 & 7    \\
1091-2 & 1091          & 52.56                         & 40.53                         & 152.81                     & -12.95                        & 4772                                 & 5    \\
1093-1 & 1093          & 241.24                        & 17.57                         & 31.31                      & 44.52                         & 10395                                & 17   \\
1093-2 & 1093          & 241.48                        & 18.00                          & 31.99                      & 44.46                         & 11589                                & 11   \\
1100-1 & 1100          & 251.94                        & 58.81                         & 88.16                      & 38.82                         & 5256                                 & 3    \\
1100-2 & 1100          & 249.56                        & 57.86                         & 87.30                       & 40.25                         & 5349                                 & 6    \\
1154-1 & 1154          & 52.59                         & 41.79                         & 152.08                     & -11.92                        & 4325                                 & 3    \\
1154-2 & 1154          & 52.13                         & 40.29                         & 152.68                     & -13.34                        & 4107                                 & 3    \\
1186-1 & 1186          & 195.68                        & -56.21                        & 304.51                     & 6.63                          & 6225                                 & 4    \\
1186-2 & 1186          & 196.91                        & -57.33                        & 305.13                     & 5.47                          & 6030                                 & 13   \\
119-1  & 119           & 142.76                        & -61.89                        & 281.56                     & -7.66                         & 2862                                 & 17   \\
119-2  & 119           & 137.59                        & -64.03                        & 281.46                     & -10.82                        & 2076                                 & 4    \\
1201-1 & 1201          & 173.22                        & -9.62                         & 272.85                     & 48.61                         & 6630                                 & 21   \\
1201-2 & 1201          & 174.68                        & -9.32                         & 274.62                     & 49.52                         & 6155                                 & 4   \\ \hline
\label{tbl:SubGroups}
\end{tabular}
\end{table*}

\begin{table*}
\caption{2MRS Group Member catalogue}
\begin{tabular}{l d{5} d{5} d{5} d{5} rrr}
\hline \hline \\
2MASS ID         & \multicolumn{1}{c}{$\alpha$}  & \multicolumn{1}{c}{$\delta$} & \multicolumn{1}{c}{$l$}       & \multicolumn{1}{c}{$b$}       &  \multicolumn{1}{c}{$v_{\rm cmb}$}        & Group ID & Sub ID \\ \\
                 & \multicolumn{4}{c}{{[}deg, J2000{]}} &  \multicolumn{1}{c}{{[}km~s$^{-1}${]}} &          &        \\  \\ \hline \hline
$02284905+3810005$ & 37.20436                      & 38.16689                      & 143.22255                     & -20.83781                     & 11189                                 & 1541     & 1541-2   \\
$10032864-1530044$ & 150.86940                      & -15.50126                    & 254.13606                     & 31.03521                      & 9998                                  & 1763     & -        \\
$04375557-0931092$ & 69.48154                      & -9.51926                      & 205.99341                     & -33.93850                      & 5110                                  & 724      & -        \\
$02362379+3142410$ & 39.09909                      & 31.71140                       & 147.68501                     & -26.06654                     & 4705                                  & 152      & 152-2    \\
$16175726-6055229$ & 244.48875                     & -60.92303                     & 325.53091                     & -7.47124                      & 3597                                  & 153      & 153-3    \\
$08333766+5535322$ & 128.40689                     & 55.59229                      & 162.22017                     & 36.37661                      & 11300                                 & 1971     & -        \\
$01293373+1717532$ & 22.39054                      & 17.29812                      & 135.76518                     & -44.62148                     & 12761                                 & 2535     & -        \\
$14165292+1048264$ & 214.22060                      & 10.80737                      & 357.96130                      & 64.11185                      & 7650                                  & 334      & -        \\
$10081231+0958374$ & 152.05138                     & 9.97701                       & 229.02748                     & 47.94034                      & 8532                                  & 2188     & -        \\
$02442183+3121169$ & 41.09096                      & 31.35475                      & 149.54651                     & -25.61624                     & 5203                                  & 152      & 152-1    \\
$14034273-3243006$ & 210.92809                     & -32.71685                     & 320.07907                     & 27.73697                      & 4028                                  & 85       & -        \\
$09202342+5456313$ & 140.09766                     & 54.94210                       & 161.67014                     & 43.04770                       & 13827                                 & 2358     & -        \\
$17281491-6640152$ & 262.06204                     & -66.67094                     & 325.71674                     & -17.07140                      & 13589                                 & 193      & -        \\
$23000358+1558493$ & 345.01495                     & 15.98034                      & 87.56570                       & -39.12365                     & 1828                                  & 354      & -        \\
$13295512-3119561$ & 202.47968                     & -31.33226                     & 312.50024                     & 30.82352                      & 15142                                 & 1782     & -        \\
$21073236-2538350$ & 316.88477                     & -25.64310                      & 21.20206                      & -40.26195                     & 11998                                 & 818      & -        \\
$12492664-4127463$ & 192.36107                     & -41.46286                     & 302.53073                     & 21.40733                      & 5086                                  & 50       & 50-1     \\
$13331326+3306350$ & 203.30525                     & 33.10971                      & 68.98088                      & 79.17372                      & 7636                                  & 149      & -        \\
$17030344+6102381$ & 255.76439                     & 61.04391                      & 90.49467                      & 36.52942                      & 3391                                  & 839      & -        \\
$19324667-6431251$ & 293.19467                     & -64.52367                     & 331.77798                     & -28.70660                      & 4233                                  & 806      & -        \\
$02094273-1011016$ & 32.42808                      & -10.18383                     & 174.08235                     & -64.95735                     & 3623                                  & 413      & -        \\
$05393634+1259047$ & 84.90136                      & 12.98459                      & 192.95486                     & -9.49321                      & 7229                                  & 2956     & -        \\
$13170188-1046121$ & 199.25781                     & -10.76999                     & 313.08246                     & 51.59659                      & 3153                                  & 1447     & -        \\
$18492430-4846088$ & 282.35126                     & -48.76904                     & 347.37582                     & -19.73818                     & 5244                                  & 2405     & -        \\
$22120645+3720024$ & 333.02695                     & 37.33405                      & 91.02950                       & -15.47828                     & 5632                                  & 189      & 189-2    \\
$03053667+0459396$ & 46.40281                      & 4.99440                        & 173.28879                     & -44.36787                     & 8407                                  & 452      & -        \\
$10512393+2806429$ & 162.84979                     & 28.11196                      & 203.77824                     & 63.45928                      & 1562                                  & 124      & -        \\
$06254878-5359361$ & 96.45339                      & -53.99345                     & 262.68002                     & -25.29278                     & 14006                                 & 2636     & -        \\
$09045016+1333424$ & 136.20901                     & 13.56185                      & 215.39915                     & 35.63015                      & 8778                                  & 1879     & -        \\
$02303366+3210342$ & 37.64031                      & 32.17624                      & 146.21729                     & -26.17465                     & 4552                                  & 593      & -        \\ \hline
\end{tabular}
\label{tbl:groupmembers}
\end{table*}

\begin{figure*}
    \centering
    \includegraphics[width=\textwidth]{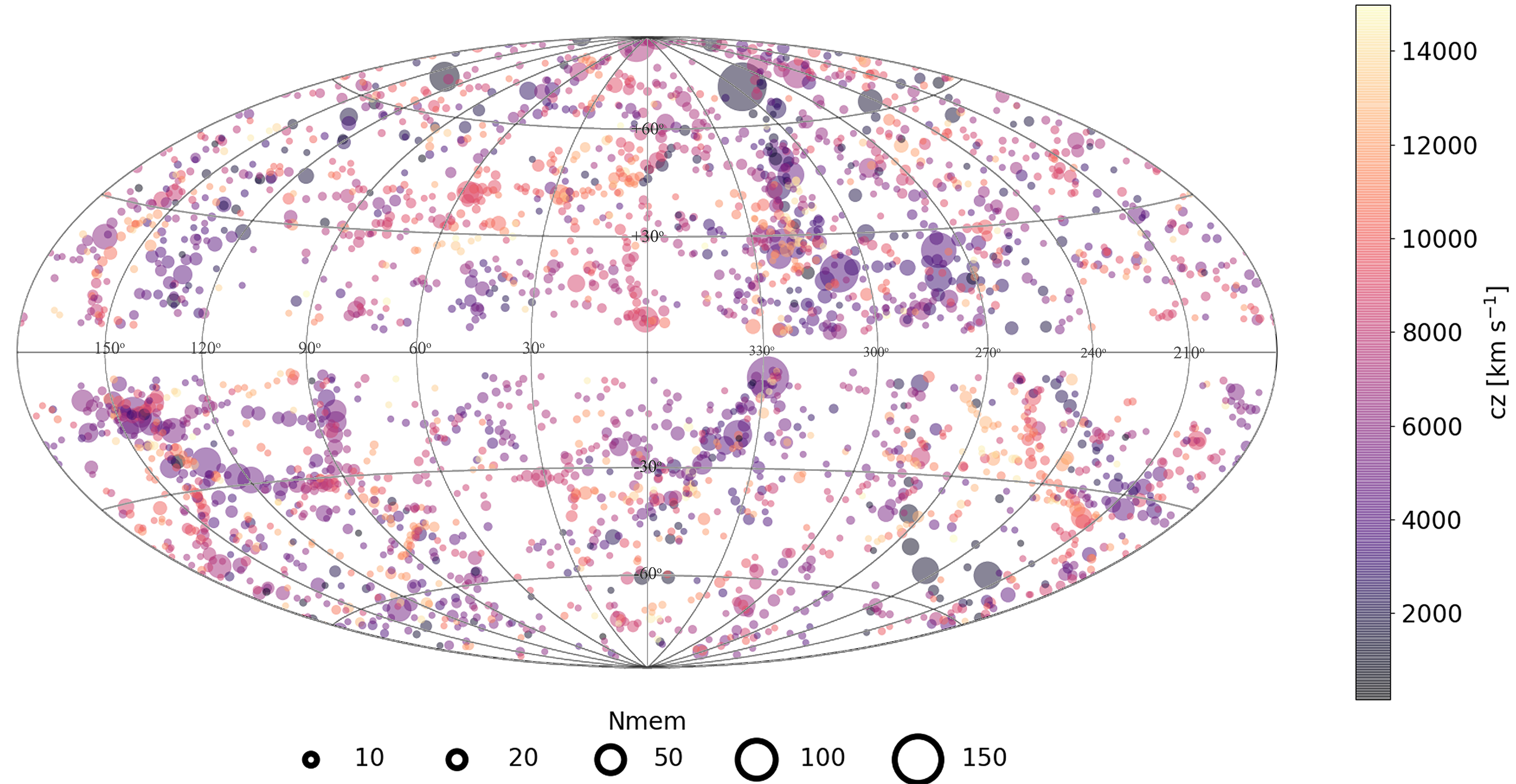}
    \caption{Aitoff projection in Galactic coordinates of the completed 2MRS group catalogue. Each symbol represents a group, with its size determined by the number of members and its color related to the recession velocity. Only groups out to $cz=15000$ km~s$^{-1}$ are shown.}
    \label{fig:FamousAitoffGroup}
\end{figure*}

We recovered all groups detected by \cite{Huchra2007}. There are several groups consisting only of galaxies with $K_{\rm s}^{\rm o}>11 \fm 25$ which are present in our catalogue but naturally are not in \cite{Huchra2007}. Interestingly, several groups in \cite{Huchra2007} were identified at the lower and upper velocity ranges of large clusters. In other cases, large clusters (such as Virgo and Coma) that were made up of numerous groups in \cite{Huchra2007} only contained one single large cluster in our catalogue. This is an example of ``shredding" that happens with large groups and clusters. Our method is robust against shredding, and hence this is one of the common differences we see in comparing with other catalogues, in this case \cite{Huchra2007}.

\subsubsection{Tempel et al., (2016 and 2018)}
\cite{Tempel2016} created a modified version of the FoF algorithm in which membership refinement was included, making use of: (1) multimodality analysis (in an attempt to identify substructure within groups) and (2) determining the virial radius of groups and identifying any outliers from the escape velocity of the members (in order to not unjustifiably add members to groups). These techniques have direct analogies within our own method to achieve many of the same goals. 

Their modified version of the FoF was applied to the \citet{Huchra2012} version of 2MRS, as well as to other redshift surveys. They compared the resulting 2MRS group catalogue with the \cite{Tully2015} 2MRS group catalogue between 3000 and 10000 km~s$^{-1}$. 

\cite{Tempel2016} and \cite{Tempel2018} allowed groups with just two members, instead of our stronger requirement of 3 or more. We applied our criterion to their results to enable a consistent comparison with our work.

The most significant difference between the \cite{Tempel2016} and \cite{Tempel2018} catalogues are the methods used to identify groups. Whilst \cite{Tempel2016} relied on a modified FoF algorithm, \cite{Tempel2018} made use of a Bayesian group finder {based on the method discussed in \cite{Stoica2010}}. Once again, their algorithm was applied to the \cite{Huchra2012} catalogue.

As was the case in \cite{Huchra2007}, both \cite{Tempel2016} and \cite{Tempel2018} had incidents in which large and well-known clusters were separated into numerous groups, whereas these are identified as a single cluster by our procedures. This further validates the success of our graph theory implementation. Groups found by \cite{Tempel2016} or \cite{Tempel2018} and not found by us generally contained very few members and were often marginally connected. We attribute these cases \cite[in][]{Tempel2016} to a larger set of linking lengths which allowed for more marginal detections. Despite this, we do recover more groups in total than either \cite{Tempel2016} or \cite{Tempel2018} as seen in the distributions in Fig. \ref{fig:CompHist}. 

\subsubsection{Lim et al., (2017)}
\cite{Lim2017} has developed a halo based group finder which was applied to 4 major redshift surveys, most notably the 2MRS. This halo-based finder provides a uniquely different method to both the FoF algorithm used by \cite{Huchra2007} and \cite{Tempel2016} and the halo-based group finder of \cite{Tempel2018}. This makes it a suitable candidate to compare to our own catalogue. 

\cite{Lim2017} included groups with members $\geq$ 1. In order to make a fair comparison we only include groups with $N \geq 3$. 

This catalogue was the closest to our own. However, there was severe shredding present within many of the large, well-known clusters and we still recovered more groups than \cite{Lim2017}. 

Of particular interest is the case of Abell 3558 (Shapley), our method found multiple groups instead of one large cluster where as \citep{Lim2017} correctly identify the cluster. This was due to the incompleteness of 2MRS at these large distances.  

\subsubsection{New Groups}
\begin{figure}
    \centering
    \includegraphics[width=0.5\textwidth]{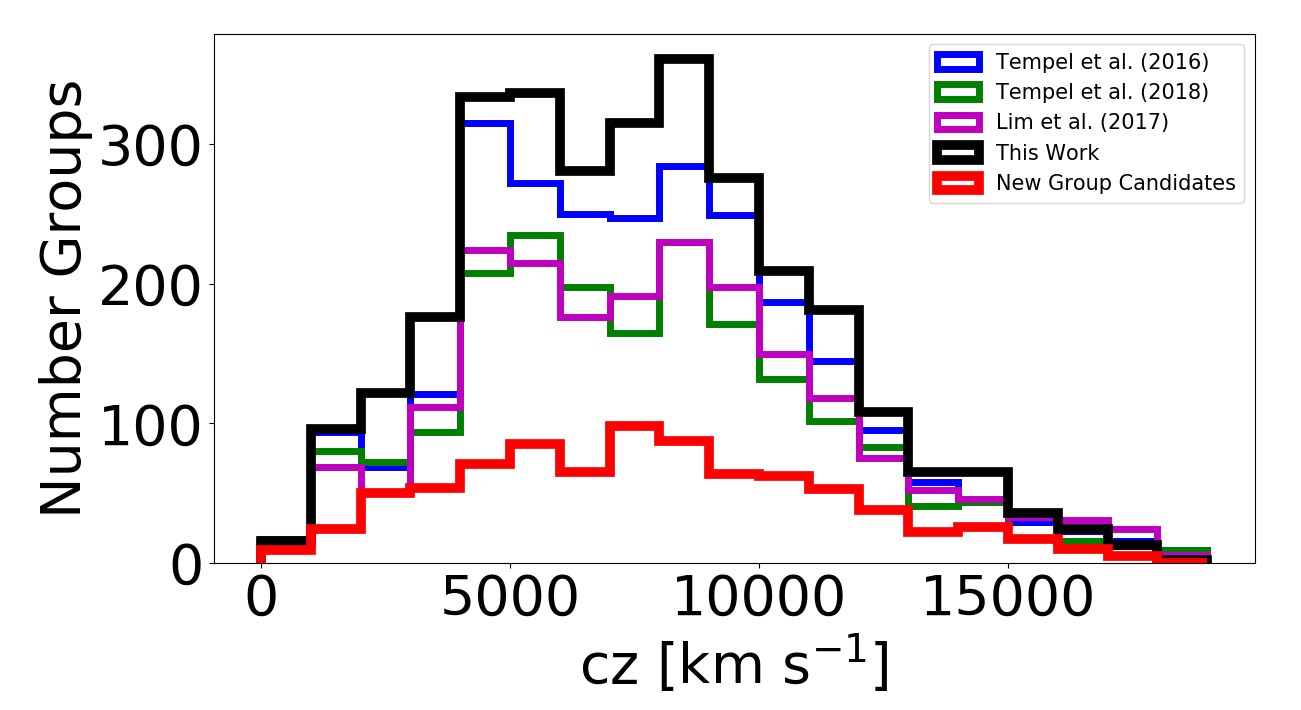}
    \caption{Number of groups as a function of redshift in our catalogue and previous analyses of 2MRS.}
    \label{fig:CompHist}
\end{figure}

\begin{figure*}
    \centering
    \includegraphics[width=\textwidth]{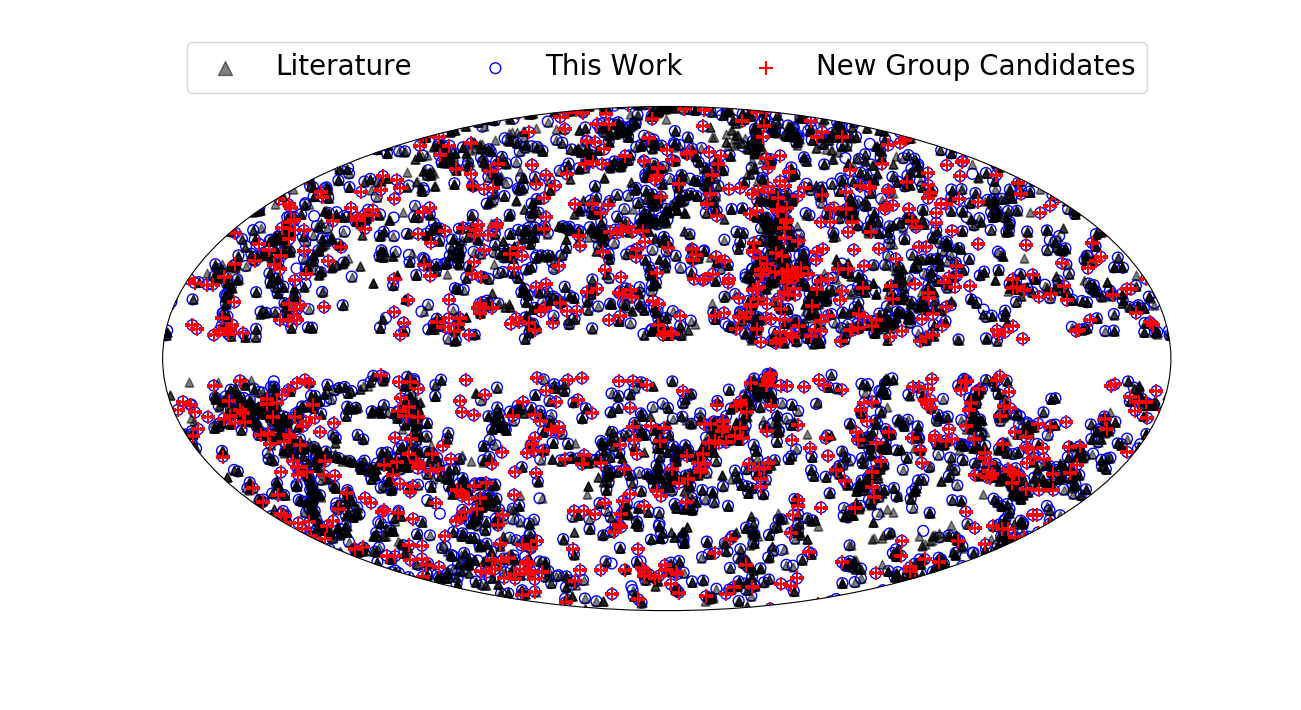}
    \caption{Galactic Aitoff projection showing the on-sky distribution of galaxy groups from \citet{Huchra2007,Lim2017,Tempel2016,Tempel2018} as well as the groups found in this work. Black triangles show literature groups, blue circles show groups from this work, and red crosses demarcate possible new groups not found in literature.}
    \label{fig:CompAitoff}

    \includegraphics[width=\textwidth]{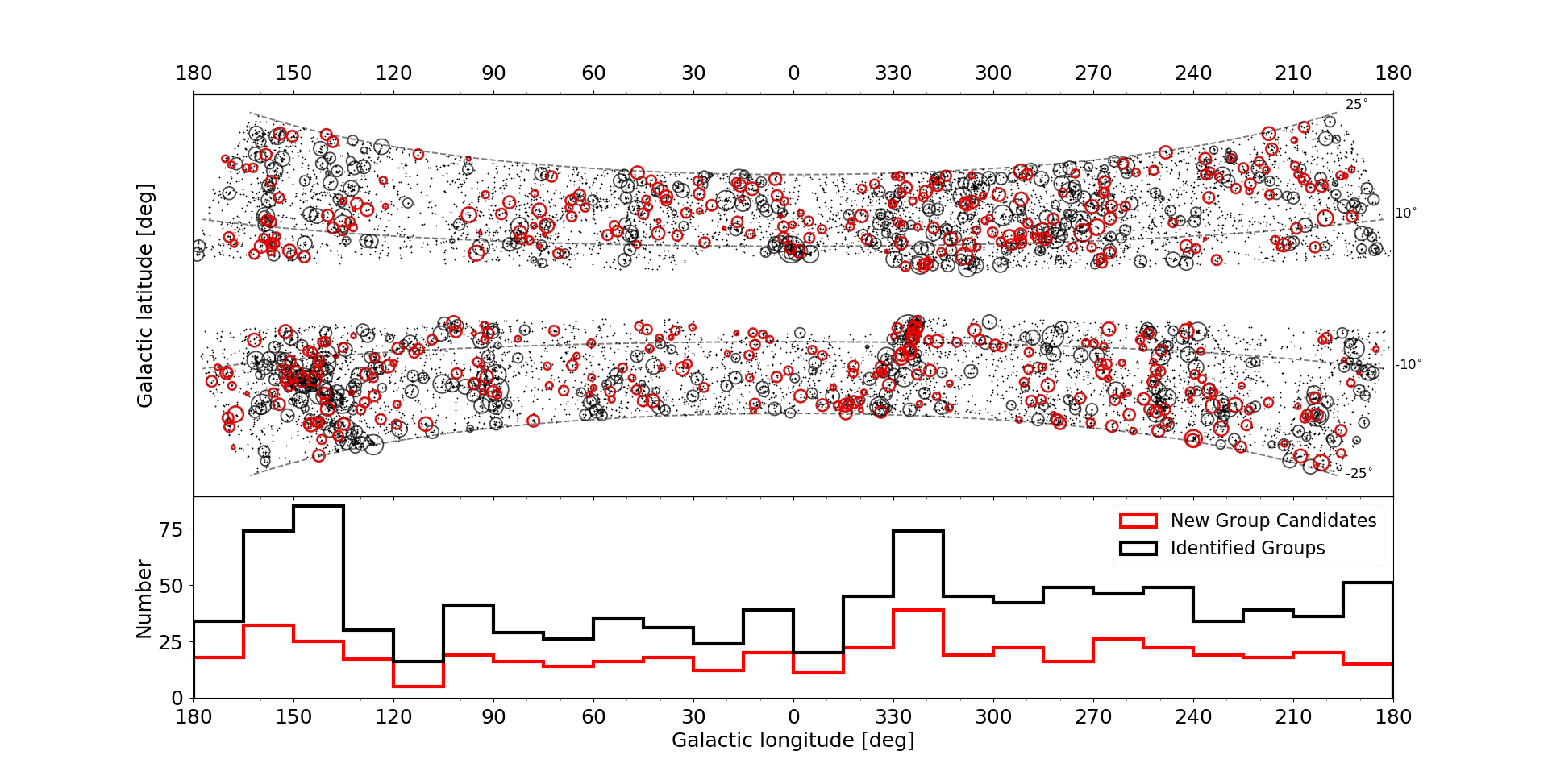}
    \caption{Zoomed-in view of the band around the ZoA ($|b| < 25^{\circ}$). Top panel shows the on-sky distribution of galaxies and galaxy groups found in the ZoA. Circles represent groups found in this work. Black circles have a matching group in the literature. Red circles are groups without a match and are new group candidates. Bottom panel shows the distribution of the groups along the Galactic plane. Black line shows the distribution of all the groups in the 2MRS while the red shows the distribution of the new groups only. }
    \label{fig:ZoomCompAitoff}
\end{figure*}

New groups were identified as those with no match (within our tolerances) in the aforementioned catalogues. Figure~\ref{fig:CompHist} shows their distribution as a function of redshift. Notably, their distribution is identical to the other catalogues, implying that new groups are being found around existing large-scale structures. This can be further verified from their spatial distribution, as seen in Fig.~\ref{fig:CompAitoff}. For the most part it seems that the new galaxy group candidates are aligned with already well-known structures; however, there are several new ones that are found in under-dense regions.

Of particular interest is the ZoA, where the majority of the new redshifts in the final release of 2MRS were located. Figure~\ref{fig:ZoomCompAitoff} shows a zoomed-in view of the ZoA ($|b|<25^{\circ}$). Once again, there are several new galaxy group candidates in regions of higher density as well as several along the ``lip'' of the exclusion Zone. These particular groups will be interesting to follow up, as many align themselves with known structures deep within the ZoA \citep[$|b|<10^\circ$,][]{KK2016Parkes,Ramatsoku2016,Courtois2019}. 

As an additional excercise in examining the validity of our new galaxy group candidates, an extreme set of cross-matching criteria was chosen (6 and 12 Mpc in the plane-of-sky and along the line-of-sight, respectively) in order to identify the most likely new galaxy group candidates. Among these we selected 20 groups with 5 or more members, only five of which were previously identified in the NASA Extragalactic Database (NED); two are substructures of the well-known Shapley cluster \citep{Zabludoff1993,Ramella1997,Bardelli1998,Mahdavi2000,Ramella2002,Ragone2006,Gimenez2015} while the other three were found in the Northern CfA redshift survey \cite{Ramella2002}. 

75\% of these new group candidates had no counterpart in NED and thus make for reasonable follow-up candidates. This illustrates that several new groups exist within the catalogue but more work would be needed to identify and verify them as such. The new 2MRS groups that were previously identified in the literature bode well for the methods used in this work, as substructure within the Shapley Supercluster could already be identified from this rather shallow survey, unlike previous analyses of 2MRS.

We include a catalogue of new galaxy group candidates based on our comparisons with the aforementioned previous work. A small sub-sample is shown in Table \ref{tbl:NewGroup_catalogue} and include Group ID, RA, Dec, $l$, $b$, $v_{\text{cmb}}$, $N_{\text{mem}}$, and $N'_{\text{mem}}$ as described in Table \ref{tbl:catalogue}.  We include candidate groups with 3 or more members but would recommend only considering those with 4 or more members as more reliable groups.  

\begin{table*}
\caption{2MRS New Galaxy Group catalogue}
\begin{tabular}{rd{2} d{2} d{2} d{2} rcc}
\hline \hline 
\\
Group ID & \multicolumn{1}{c}{RA J2000}  & \multicolumn{1}{c}{Dec J2000} & \multicolumn{1}{c}{$l$}       & \multicolumn{1}{c}{$b$}       & \multicolumn{1}{c}{$v_{ \rm cmb}$}           & $N_{\rm mem}$ & $N'_{\rm mem}$ \\ \\
         & \multicolumn{1}{c}{{[}deg{]}} & \multicolumn{1}{c}{{[}deg{]}} & \multicolumn{1}{c}{{[}deg{]}} & \multicolumn{1}{c}{{[}deg{]}} & \multicolumn{1}{c}{{[}km~s$^{-1}${]}} &      &   \\ \\ 
(1) & \multicolumn{1}{c}{(2)}  & \multicolumn{1}{c}{(3)} & \multicolumn{1}{c}{(4)}       & \multicolumn{1}{c}{(5)}       & \multicolumn{1}{c}{(6)}           & (7) & (8)  \\ \\         
         \hline \hline
1782     & 202.17                        & -31.52                        & 285.13                        & 73.58                         & 15560                                 & 14   & 285        \\
2051     & 87.07                         & -25.58                        & 325.27                        & -7.17                         & 11479                                 & 12   & 130       \\
2186     & 54.99                         & 42.49                         & 302.26                        & 21.66                         & 9894                                  & 8    & 54        \\
2150     & 86.74                         & -25.60                         & 150.53                        & -13.45                        & 12844                                 & 8    & 100    \\
1579     & 267.92                        & 7.70                          & 59.10                         & 87.99                         & 6349                                  & 7    & 22        \\
576      & 89.65                         & -52.36                        & 269.64                        & 26.35                         & 9901                                  & 7    & 48        \\
1435     & 254.18                        & -6.31                         & 285.13                        & 73.58                         & 8699                                  & 6    & 31         \\
1692     & 193.89                        & -11.85                        & 304.14                        & 49.10                         & 6591                                  & 6    & 20         \\
2339     & 54.96                         & -2.56                         & 317.13                        & 30.92                         & 10392                                 & 6    & 47         \\
2307     & 240.56                        & -62.21                        & 136.75                        & -24.96                        & 13528                                 & 6    & 75        \\
225      & 325.56                        & -70.96                        & 332.53                        & -23.38                        & 3686                                  & 6    & 10        \\
2645     & 245.84                        & 39.89                         & 234.45                        & 73.12                         & 9679                                  & 6    & 38         \\
439      & 220.02                        & -37.06                        & 236.00                        & -54.21                        & 4559                                  & 6    & 13         \\
1844     & 338.07                        & 51.93                         & 273.17                        & 19.27                         & 11315                                 & 6    & 65         \\
1704     & 207.24                        & -50.68                        & 59.10                         & 87.99                         & 8586                                  & 6    & 31         \\
2374     & 106.14                        & 53.95                         & 142.70                        & -63.07                        & 11190                                 & 6    & 61         \\
1613     & 170.55                        & -1.11                         & 130.51                        & -28.86                        & 7878                                  & 6    & 26         \\
1984     & 240.53                        & 36.70                         & 325.27                        & -7.17                         & 9364                                  & 6    & 36        \\
97       & 116.21                        & -51.12                        & 145.21                        & 68.92                         & 1217                                  & 6    & 6         \\
1628     & 155.34                        & -4.57                         & 150.53                        & -13.45                        & 12209                                 & 6    & 75        \\
1157     & 39.55                         & 35.39                         & 167.35                        & 22.93                         & 9166                                  & 6    & 36         \\
1508     & 91.52                         & -35.81                        & 140.83                        & -17.36                        & 9824                                  & 6    & 41        \\
2522     & 256.37                        & 25.09                         & 302.26                        & 21.66                         & 11465                                 & 6    & 65         \\
338      & 12.60                         & -2.13                         & 319.22                        & 26.81                         & 3660                                  & 6    & 10         \\
2597     & 318.00                         & -48.49                        & 8.70                          & -27.09                        & 9333                                  & 5    & 30         \\ \hline
\end{tabular}
\label{tbl:NewGroup_catalogue}
\end{table*}

\section{Discussion}
Fig.~\ref{fig:FamousAitoffGroup} shows the distribution of galaxy groups from Table \ref{tbl:catalogue} in Galactic coordinates, with the size and color of the symbols related to the number of members ($N_{\text{mem}}$) and recession velocity, respectively. All the major well-known large-scale structures have been recovered, including: the Virgo Supercluster, the Coma cluster, the Perseus-Pisces complex, the Fornax cluster, the Norma cluster (Great Attractor), and the Ophiuchus cluster, to name a few. This bodes well for validating our method, as these clusters were all identified in many previous analyses of 2MRS \citep{Jarrett2004,Strutskie2006,Huchra2007,Huchra2012,Macri2019}. Even more interesting is examining the new groups within the ZoA, shown in Fig.~\ref{fig:ZoomCompAitoff}. 

Several structures within the ZoA were mentioned in \cite{Macri2019}, such as the Perseus-Pisces complex and its two extensions through the ZoA at $l \approx 90^{\circ}$ and $165^{\circ}$. In Fig.~\ref{fig:ZoomCompAitoff} we can indeed confirm that the galaxies not only map out an extension of the Perseus-Pisces chain from $l \approx 150 ^{\circ}$ to $l \approx 160^{\circ}$ \citep[as discussed in][]{Macri2019} but so do the galaxy groups which later connect to Lynx at $l \approx 170^{\circ}$ \citep{Ramatsoku2016,KK2018,Macri2019}. There are also several new group candidates at this crossing, further emphasising the importance of filling in the ZoA. The second crossing of the Perseus-Pisces chain, at $l\approx 90 ^{\circ}$, also includes several new groups and is well defined in our group catalogue. \cite{Macri2019} also highlights a surprising new density at $(l,b)\approx (100^{\circ},-5^{\circ}$). In Figure \ref{fig:ZoomCompAitoff} we see that there are groups in this over-density, several of which have been identified as new and therefore may be worth additional follow up observations.  

\section{Conclusions and Next Steps}
The traditional FoF algorithm developed by \cite{Huchra1982} has been a reputable and well-validated method of identifying galaxy groups in redshift surveys. However, it has several shortcomings mostly arising from a static set of linking lengths. This leads to large clusters being identified as several smaller groups or several small groups falsely being identified as a single large cluster. Several modifications of the algorithm over the years attempted to address some of these issues: from using two different sets of linking lengths and reporting both low and high density catalogues \citep{Huchra2007}, to adopting membership refinement based on escape velocities \citep{Tempel2016}. 

We have developed our own, graph-theory based, modification to the traditional FoF algorithm in order to address most shortcomings. Our group finder: {(1) reliably identifies large clusters as a single body without identifying smaller non-physical groups as a single cluster, (2) reliably identifies small groups without non-physical pieces originating from shredded larger clusters, (3) is robust to outliers and chance alignments of galaxies, and (4) provides a unique method of identifying substructures within larger groups and clusters.} However, while our method provides improvements over the traditional FoF algorithm, we are always susceptible to z-axis confusion and chance alignments of separate galaxies along the radial direction. This is due to the redshift distortion present in all redshift surveys and is something that FoF algorithms have a difficult time with in general.

Our group finder was run on the recently-completed 2MRS \citep{Macri2019} and as such differs from previous analyses as being the deepest to date and first 100$\%$ complete 2MRS galaxy group catalogue to a magnitude limit of $K_{s}^o \leq 11\fm75$. Comparisons to previous work show that this method is able to recover most, if not all, the groups of previous catalogues and is able to identify numerous more, including substructures of the Shapley Supercluster for the first time. As such we have included a catalogue of new, previously-unidentified group candidates which might be entirely new large-scale structures. These provide an interesting follow up opportunity, as additional observations are required to validate them.

This novel adaptation lent itself to several visual techniques, with both our method and final galaxy group catalogue being extensively and exhaustively interrogated using the new IDIA visualization laboratory -- including immersive screen technology, virtual reality, and the newly upgraded Iziko digital dome planetarium. 

This group finder is currently in the process of being generalized in order to be run on any other magnitude-limited redshift surveys. It will be released as a Python package, allowing users to quickly and reliably generate their own group catalogues. {The current version of the group finder is available at \url{https://github.com/BrutishGuy/pyfriends}}

While our 2MRS galaxy group catalogue is the most complete, ``whole-sky'' work to date, there still remains a large gap within the ZoA which \cite{Macri2019} necessarily excluded. Recently, \cite{Schroder2019} has created a catalogue of bright 2MRS galaxies within this area; we are currently obtaining their redshifts to be added to a final 2MRS galaxy group catalogue which will be applicable to studies of large-scale structures which are near or cross the Zone of Avoidance. 

\section*{Acknowledgements}
This paper uses observations carried out at the South African Astronomical Observatory (SAAO), the Fred L.~Whipple Observatory (FLWO), and the SOuthern Astrophysical Research observatory (SOAR). TSL acknowledges visitor support from the Mitchell Insitute for Fundamental Physics \& Astronomy at Texas A\&M University, and is thankful for the support provided via the SARChI grants of THJ and RKK. THJ acknowledges support from the National Research Foundation, under the South Africa Research Chair Initiative. We thank Darren Croton (Swinburne) for help with the TAO mock analysis. This work is based upon research supported by the South African Research Chairs Initiative of the Department of Science and Technology and National Research Foundation. This publication makes use of data products of the Two Micron All Sky Survey, which is a joint project of the University of Massachusetts and the Infrared Processing and Analysis Center, funded by the NASA and the National Science Foundation. This work was supported in part by the Jet Propulsion Laboratory, California Institute of Technology, under a contract with NASA.

\section*{Data Availability}
All data are incorporated into this article's online supplementary material and are also available on request. 




\bibliographystyle{mnras}
\bibliography{trystansbib} 








\bsp	
\label{lastpage}
\end{document}